\documentclass{article}

\usepackage{adjustbox}
\usepackage{amsmath,amssymb,amsfonts}
\usepackage[linesnumbered,ruled,vlined]{algorithm2e}
\usepackage{setspace}
\usepackage{algorithmic}
\usepackage{float}
\usepackage{graphicx}
\usepackage{textcomp}
\usepackage{xcolor}
\usepackage{multirow}
\usepackage{mdframed}
\usepackage{balance}
\usepackage{booktabs}
\usepackage{tcolorbox}
\usepackage{makecell}
\usepackage{threeparttable}
\usepackage{colortbl}
\usepackage{subfigure}
\usepackage{hhline}
\usepackage{pifont}
\usepackage{listings}
\usepackage{enumitem}
\newcommand{\tabincell}[2]{\begin{tabular}{@{}#1@{}}#2\end{tabular}}  

\newcommand{\parabf}[1]{\noindent\textbf{#1}}

\newcommand{\detectbugnum}{10}
\newcommand{\confirmbugnum}{6}
\newcommand{\cvenum}{6}

\newcommand{\csd}{functional semantics}

\definecolor{ggray}{HTML}{eff0f0}
\definecolor{gggray}{HTML}{E8E8E8}
\definecolor{ggggray}{HTML}{BEBEBE}
\definecolor{myblue}{RGB}{255,255,255}
\definecolor{myyellow}{HTML}{FFF2CC}

\newcommand{\ie}{\textit{i.e.,}\xspace}
\newcommand{\eg}{\textit{e.g.,}\xspace}

\newcommand{\app}{\textsc{Vul-RAG}}
\newcommand{\ourbench}{PairVul}

\newcommand{\pairacc}{{pairwise accuracy}}

\usepackage[utf8]{inputenc}
\usepackage[english]{babel}

\newcounter{finding}

\newcommand{\distance}{5pt}
\usepackage{neurips_2025}

\usepackage[utf8]{inputenc} 
\usepackage[T1]{fontenc}    
\usepackage{hyperref}       
\usepackage{url}            
\usepackage{booktabs}       
\usepackage{amsfonts}       
\usepackage{nicefrac}       
\usepackage{microtype}      

\title{Vul-RAG: Enhancing LLM-based Vulnerability Detection via Knowledge-level RAG}

\author{
    Xueying Du\textsuperscript{1} \quad Geng Zheng\textsuperscript{2} \quad
    Kaixin Wang\textsuperscript{1} \quad Yi Zou\textsuperscript{1} \quad Yujia Wang\textsuperscript{1} \quad Wentai Deng\textsuperscript{1} \\
    \textbf{Jiayi Feng\textsuperscript{1} \quad Mingwei Liu\textsuperscript{3} \quad Bihuan Chen\textsuperscript{1} \quad Xin Peng\textsuperscript{1} \quad Tao Ma\textsuperscript{2} \quad Yiling Lou\textsuperscript{1}} \\
    \textsuperscript{1}Fudan University, China \\
    \textsuperscript{2}Alibaba Group, China \\
    \textsuperscript{3}Sun Yat-sen University, China}

\begin{document}

\maketitle
\begin{abstract}
Although LLMs have shown promising potential in vulnerability detection, this study reveals their limitations in distinguishing between vulnerable and similar-but-benign patched code (only 0.06 - 0.14 accuracy). It shows that LLMs struggle to capture the root causes of vulnerabilities during vulnerability detection. 
To address this challenge, we propose enhancing LLMs with multi-dimensional vulnerability knowledge distilled from historical vulnerabilities and fixes. We design a novel \textit{knowledge-level} Retrieval-Augmented Generation framework \app{}, which improves LLMs with an accuracy increase of 16\% - 24\% in identifying vulnerable and patched code. 
Additionally, vulnerability knowledge generated by \app{} can further (1) serve as high-quality explanations to improve manual detection accuracy (from 60\% to 77\%), and (2) detect \detectbugnum{} previously-unknown bugs in the recent Linux kernel release with \cvenum{} assigned CVEs. 
\end{abstract}

\section{Introduction}
\label{sec:intro}

Software vulnerabilities can cause severe consequences.
To date, there has been a large body of research on automated vulnerability detection, utilizing traditional program analysis or deep learning techniques. More recently, the advance of large language models (LLMs) further boosts learning-based vulnerability detection. Due to the strong code comprehension capabilities, LLMs show promise in analyzing malicious behaviors (e.g., detecting bugs or vulnerabilities) in code~\cite{DBLP:journals/corr/abs-2308-12697, DBLP:conf/icse/YangGMH24, DBLP:journals/corr/abs-2401-17010, li2023hitchhikers, sun2023gpt, ding2024vulnerabilitydetectioncodelanguage, widyasari2024chatgptenhancingsoftwarequality, 10.1145/3639476.3639762}.

While significant research has been dedicated to evaluating LLMs for vulnerability detection~\cite{llmvulsurvey}, their ability to accurately distinguish between vulnerable code and its corresponding patched code remains unclear. Given that vulnerable and patched code pairs often share high textual similarity, addressing this question can reveal whether LLMs genuinely capture the root causes of vulnerabilities or merely overfit to superficial code features when classifying code as vulnerable or benign. Additionally, this question is closely related to the robustness of LLMs in vulnerability detection, which reflects how well LLMs perform in distinguishing between similar code.

\parabf{Empirical Study.}
To fill this gap, we perform an empirical study to evaluate the capabilities of LLMs in distinguishing between vulnerable and patched code. We first construct a new benchmark \ourbench{}, which includes 586 high-quality pairs of vulnerable and patched functions extracted from real-world CVEs of complicated software systems.
Our experiments reveal that existing LLMs struggle to distinguish between vulnerable and patched code: for majority (86\% - 94\%) cases, existing LLMs cannot identify the vulnerable code as vulnerable while identify its patched code as benign at the same time.
We further investigate how advanced prompts proposed in recent vulnerability detection work~\cite{10.1145/3653718,10.1145/3639476.3639762} can eliminate such limitations, including two Chain-of-Thought prompts and one CWE description enhanced prompt. 
Additionally, we explore whether fine-tuning LLMs on vulnerable-patched datasets can improve their performance in this task.
We find that all these advanced strategies bring limited improvement to LLMs, with only 0.05 - 0.20 accuracy in correctly identifying both vulnerable and patched code. 
Based on further analysis, we find that LLMs show unstable bias by dominantly identifying most code as vulnerable or benign when working with different prompts. Particularly, LLMs struggle to distinguish the subtle textual difference between vulnerable and patched code, such as relocated method invocations or modified conditional checks. \textbf{Overall, LLMs still fall short in understanding vulnerable behaviors in code.}

\parabf{Enhancement Framework \app{}.} 
To address this challenge, we propose \app{}, a novel knowledge-level Retrieval-Augmented Generation (RAG) framework to enhance LLM-based vulnerability detection. The key insight behind \app{} is to distill \textit{high-level, generalizable vulnerability knowledge} from historical vulnerabilities and fixes, which can guide LLMs to more accurately understand vulnerable and benign behaviors in code.
Specifically, \app{} proposes a novel multi-dimension representation (including perspectives of functional semantics, vulnerability root causes, and fixing solutions) for vulnerability knowledge. The representation focuses on high-level features of vulnerabilities rather than lexical code details.
Based on this representation, \app{} incorporates a three-step workflow for vulnerability detection. First, \app{} constructs a vulnerability knowledge base by extracting multi-dimension knowledge from existing CVE instances and fixes via LLMs; Second, for the given code, \app{} retrieves the relevant vulnerability knowledge with similar functional semantics; Finally, \app{} uses LLMs to assess the vulnerability of the given code by reasoning through the presence of vulnerability causes and fixing solutions from the retrieved knowledge.

\parabf{Evaluation.} 
We evaluate \app{} in extensive settings. 
(1) \textit{Evaluation on Distinguishing Capabilities.}  Our  results show that \app{} can substantially enhance the ability of various LLMs to distinguish between vulnerable and patched code (i.e., achieving 16\% -24\%  improvements in pair accuracy). Meanwhile, \app{} achieves an 9\%-14\%/7\%-11\% increase in balanced precision/recall for vulnerability detection. Our ablation study shows the superiority of our knowledge-level RAG compared to existing code-level RAG based and fine-tuning based baselines, i.e., 16\%-27\% and 22\%-26\% increase in pair accuracy.
(2) \textit{User Study on Manual Vulnerability Detection.} To evaluate the quality and usability of \app{} generated vulnerability knowledge, we conduct a user study in which participants are asked to confirm vulnerability detection results (both true positives and false alarms) with or without the assistance of \app{} generated vulnerability knowledge. The results show that the vulnerability knowledge improves manual confirmation accuracy from 60\% to 77\%. User feedback also confirms the high quality of the generated knowledge in terms of the helpfulness, preciseness, and generalizability. 
(3) \textit{Case Study on Detecting Previously-Unknown Vulnerabilities.} To evaluate whether \app{} can detect new vulnerabilities, we apply \app{} to the recent Linux kernel release (v6.9.6). \app{} detects 10 previously-unknown bugs with \cvenum{} assigned CVEs. \textbf{Our extensive evaluation shows that high-level vulnerability knowledge is a promising direction for enhancing LLM-based vulnerability detection.}

This paper makes the following contributions:
\begin{itemize}[left=0pt, topsep=1pt, itemsep=0pt, parsep=0pt]
    \item We perform the first study to reveal the limited capabilities of LLMs in differentiating vulnerable code from patched code. 

    \item We propose \app{}, a novel knowledge-level RAG framework to enhance LLM-based vulnerability detection with  generalizable and multi-dimensional vulnerability knowledge distilled from historical vulnerabilities and fixes. 

    \item We perform quantitative experiments, user study, and case analysis to extensively evaluate \app{}. The results not only show the effectiveness of \app{} in improving overall precision/recall and distinguishing capabilities of LLMs, but also show the usability of \app{} in helping manual vulnerability comprehension and detecting previously-unknown bugs for complex software (e.g., Linux Kernel). \textit{Data and code of our work are at~\cite{replication_package} with MIT license.}
    \end{itemize}

\section{Related Work}

\parabf{Empirical Studies.} 
Many efforts have been dedicated to evaluating LLMs in vulnerability detection~\cite{DBLP:journals/corr/abs-2311-16169,primevul,llmvulsurvey}, covering diverse benchmarks, LLMs, and  metrics.  
Different from existing studies, we focus on evaluating the capabilities of LLMs in distinguishing between vulnerable and patched code. Risse et al.~\cite{DBLP:conf/uss/RisseB24} evaluate such capabilities of small pre-trained models (e.g., CodeBERT, UniXcoder, and PLBart), while we study more recent instructed and large LLMs. Ullah et al.~\cite{DBLP:conf/sp/UllahHPPCS24} evaluate such capabilities of LLMs on a small sample (only 30 pairs) while we extensively study 597 pairs with both quantitative and qualitative analysis.

\parabf{Enhancing LLMs in Vulnerability Detection.}
The majority of existing work focuses on prompt engineering~\cite{10.1145/3639476.3639762, DBLP:journals/corr/abs-2312-05275}, such as chain-of-thought~\cite{wei2023chainofthought, zhang2022automatic} and few-shot learning~\cite{brown2020language}, to facilitate more powerful LLM-based vulnerability detection. Additionally, recent work explores fine-tuning approaches~\cite{DBLP:conf/icse/YangGMH24, DBLP:journals/corr/abs-2401-17010, mao2024effectivelydetectingexplainingvulnerabilities} or integration with static analysis~\cite{li2023hitchhikers, sun2023gpt, Li24Enhancing, 10.1145/3653718, li2024llmassistedstaticanalysisdetecting} to enhance LLMs in vulnerability detection. As fine-tuning enhancement often works for small models with high-quality training data and static analysis enhancement often works on specific types of bugs, in this work, we mainly focus on enhancement techniques with prompt engineering. 

\parabf{Retrieval-Augmented Generation (RAG) for Code-related Tasks.} 
RAG has been widely explored in many code-related tasks, including code generation~\cite{wang2024coderagbenchretrievalaugmentcode}, code translation ~\cite{bhattarai2024enhancing}, program repair~\cite{10.1145/3611643.3616256}, and vulnerability detection in smart contracts~\cite{Yu2024RetrievalAG}. While existing work remains on code-level RAG (retrieving and augmenting with code), \app{} is novel in using high-level, generalizable knowledge to augment generation for the source code vulnerability detection task.

\vspace{-1mm}
\section{Empirical Study}~\label{sec:study}
\vspace{-10mm}
\subsection{Experimental Setup}~\label{sec:setup}
The following RQs aim to evaluate how LLMs distinguish between vulnerable and patched code. 

\begin{itemize}[left=0pt, topsep=1pt, itemsep=2pt, parsep=2pt]

\item \textbf{RQ1:} How effectively do LLMs distinguish between vulnerable and patched code?

\item \textbf{RQ2:} How do state-of-the-art prompting strategies improve LLMs in distinguishing between vulnerable and patched code?

\end{itemize}

\subsubsection{Studied LLMs and Baselines}\label{sec:baseline}
We include four state-of-the-art LLMs that have been widely used in vulnerability detection, including two closed-source models, i.e., GPT-4o~\cite{gpt4}, Claude Sonnet 3.5~\cite{claude}, and two open-source models, i.e., Qwen2.5-Coder-32B-Instruct~\cite{qwen}, DeepSeek-V2-Instruct~\cite{deepseek}. 

In RQ1, we evaluate the capabilities of studied LLMs with a basic prompt~\cite{DBLP:conf/issre/PurbaGRC23}.
In RQ2, we investigate three state-of-the-art prompting strategies proposed in recent LLM-based vulnerability detection work~\cite{10.1145/3653718,10.1145/3639476.3639762}. These include (1) two prompts that combine role-oriented with chain-of-thought, one involving an initial explanation of code behavior, and the other focusing on the root causes reasoning of vulnerabilities (denoted Cot-1 and Cot-2); 
and (2) a prompt enhanced with CWE descriptions (denoted CWE-enhanced).
The detailed prompts and baseline settings  are in Appendix~\ref{ap:baseline}.

\subsubsection{Benchmark}
\label{sec:benchmark}
Existing widely-used vulnerability detection benchmarks, such as  BigVul~\cite{DBLP:conf/msr/FanL0N20}, Devign~\cite{DBLP:conf/nips/ZhouLSD019} and Reveal~\cite{DBLP:journals/tse/ChakrabortyKDR22} are not directly applicable for our study, due to (1) the lack of corresponding patched versions for vulnerable code (e.g., Devign and Reveal), and (2) the absence of verified correctness for patched code. For example, although BigVul includes patched code, its patches may have been subsequently modified in later CVEs, making their correctness unreliable. 
Therefore, we construct a new benchmark \ourbench{}, which specifically targets high-quality pairs of vulnerable functions and their corresponding patched functions. 

Our benchmark construction process includes three key steps. (1) \textit{Vulnerable and Patched Code Collection}: We extract function-level pairs of vulnerable and patched code, along with descriptions from existing CVEs of real-world systems (i.e., Linux kernel). Particularly, we focus on Top-10 prevalent CWEs (i.e., 416, 476, 362, 119, 787, 20, 200, 125, 264, 401).  (2) \textit{Patched Code Verification}: To ensure the reliability of the patched code, we manually summarize multiple filtering rules to verify the patched code is not subsequently reverted/modified by other commits. (3) \textit{Pair Selection:} To ensure the diversity of the benchmark and control the benchmark scale, we randomly sample one-fourth of the vulnerable-patched function pairs from each CWE to construct the final benchmark. We exclude cases where the code length exceeds the current token limit of studied LLMs (i.e., 16,384 tokens). 
As a result, \ourbench{} includes 586 pairs across 420 CVEs.
Detailed construction procedure and benchmark statistics are in Appendix~\ref{ap:dataset}.

\subsubsection{Metrics}
We focus on the following metrics. \textbf{\pairacc{}} calculates among all pairs, the ratio of pairs whose vulnerable and patched code are both correctly identified. We use \textbf{Balanced Recall} (defined as $\left(\frac{\text{\#True}_{\text{vul}}}{\text{\#Total}_{\text{vul}}} + \frac{\text{\#True}_{\text{nvul}}}{\text{\#Total}_{\text{nvul}}}\right) / 2$) and \textbf{Balanced Precision}  (defined as $\left(\frac{\text{\#True}_{\text{vul}}}{\text{\#Predict}_{\text{vul}}} + \frac{\text{\#True}_{\text{nvul}}}{\text{\#Predict}_{\text{nvul}}}\right) / 2$) to evaluate the precision and recall across both vulnerable and non-vulnerable instances. Notably, Balanced Recall is equivalent to the overall accuracy given the even distribution of vulnerable and non-vulnerable samples on \ourbench{}.

\begin{table}[htb]
    \centering
    \caption{Evaluation of Basic LLMs}\label{table:rq1}
    \footnotesize
    \begin{adjustbox}{width=0.5\linewidth}
        \begin{tabular}{c|c|c|c}
        \hline

        \textbf{LLMs} & \textbf{Pair Acc.} & \textbf{Bal. Recall} & \textbf{Bal. Pre.} \\ \hline
        GPT-4o & 0.08  & 0.50 & 0.49\\
        Claude & 0.06  & 0.50 & 0.50\\
        Qwen & 0.07  & 0.52 & 0.52\\
        DeepSeek & 0.14 & 0.51 & 0.52\\
        
        \hline
        \end{tabular}
    \end{adjustbox}
\end{table}
\vspace{-2mm}
\subsection{RQ1: Basic Differentiating Capabilities}\label{sec:study_rq1}
Table~\ref{table:rq1} presents the effectiveness of LLMs with the  basic prompt. All LLMs show limited capabilities of distinguishing vulnerable and patched code.  The low \pairacc{} (i.e., 0.06 - 0.14) show that LLMs fail to accurately identify a pair of vulnerable and patched code for majority cases (86\% - 94\%). Additionally, all LLMs show limited balanced recall and precision (not more than 0.52), which is similar as random guess.

\subsection{RQ2: Impact of Advanced Prompting}\label{sec:study_rq2}
Table~\ref{table:rq2} shows that the advanced prompts bring limited improvements on the distinguishing capabilities of LLMs. Even for the best case (CoT-1 for GPT-4o), its \pairacc{} is only improved to 0.20, while others bring fewer improvements and some (CWE-enhanced) even harm the pair accuracy. 
Additionally, the balanced precision and accuracy still remain limited (lower than 0.55). 

\begin{table}[htb]
    \centering
    \caption{Impacts of Enhancement Techniques}\label{table:rq2}
    \footnotesize
    \begin{adjustbox}{width=0.55\linewidth}
        \begin{tabular}{c|c|c|c|c}
        \hline
        \textbf{Tech.} & \textbf{LLM} & \textbf{Pair Acc.} & \textbf{Bal. Recall} & \textbf{Bal. Pre.}\\ \hline

        \multirow{4}{*}{CoT-1}
        & GPT-4o   & \textbf{0.20}  & 0.52  & 0.52  \\
        & Claude   & 0.13  & 0.51  & 0.51\\
        & Qwen     & 0.09  & 0.50  & 0.50 \\
        & DeepSeek & 0.17  & 0.53  & 0.53 \\\hline 
        
        \multirow{4}{*}{CoT-2}
        & GPT-4o   & 0.18  & 0.50  & 0.50 \\
        & Claude   & 0.12  & 0.51  & 0.51  \\
        & Qwen     & 0.17  & 0.52  & 0.52 \\
        & DeepSeek & 0.15  & 0.53  & 0.54\\\hline 

        \multirow{4}{*}{\tabincell{c}{CWE\\Enhanced}}
        & GPT-4o   & 0.13  & 0.51  & 0.51 \\
        & Claude   & 0.12  & 0.53  & 0.53  \\
        & Qwen     & 0.05  & 0.51  & 0.51 \\
        & DeepSeek & 0.18  & \textbf{0.55}  & \textbf{0.55} \\ \hline

        \end{tabular}
    \end{adjustbox}
\end{table}

\begin{table}[htb]
    \centering
    \caption{Vulnerable Code Identification Ratio}
    \label{table:rq2_rv}
    \footnotesize
    \begin{adjustbox}{width=0.55\linewidth}
        \begin{tabular}{c|c|c|c|c}
        \hline
        \textbf{Technique} & \textbf{GPT-4o} & \textbf{Claude} & \textbf{Qwen} & \textbf{DeepSeek} \\
        
        \hline
        Basic LLMs   & 0.77 & 0.49 & 0.71 & 0.81 \\
        CoT-1        & 0.48 & 0.34 & 0.17 & 0.72 \\
        CoT-2        & 0.67 & 0.66 & 0.61 & 0.80 \\
        CWE-Enhanced & 0.36 & 0.34 & 0.20 & 0.40 \\

        \hline
        \end{tabular}
    \end{adjustbox}
\end{table}

\parabf{Further Analysis.} 
We perform quantitative and qualitative analysis of RQ1 and RQ2 results.

\textit{Unstable Bias.} Table~\ref{table:rq2_rv} shows the ratio of cases that LLMs identify the code as vulnerable when working with different prompts. Interestingly, we find that LLMs show unstable biases although all the prompts are neutral without inductive instructions. With a basic prompt, GPT-4 and DeepSeek tend to consider majority code (over 75\%) as vulnerable. The CoT-1 (explaining the code behaviors first) and CWE-enhanced (including the relevant CWE descriptions) dramatically lead most LLMs (except DeepSeek) to consider most code as benign. The results further confirm that LLMs cannot capture the semantic difference between vulnerable and patched code, thus showing unstable bias when instructed with different neutral prompts. 

\textit{Case Analysis.} We manually sample and analyze code pairs where all the studied LLMs fail to distinguish between vulnerable and patched code. We further confirm that it is challenging for LLMs to discern the subtle textual differences between two similar functions with opposing labels (i.e., vulnerable vs. benign), such as (1) relocating a method invocation, (2) replacing a method invocation, and (3) adding a conditional check. Detailed bad case examples are in Appendix~\ref{ap:case_study_rq}. 

\parabf{Summary of Empirical Studies.} Overall, our empirical study reveals that LLMs cannot distinguish between vulnerable  and patched code (i.e., pair accuracy lower than 0.14 and balanced recall/precision lower than 0.52), while the recent prompting techniques bring limited improvements. LLMs show unstable bias with different neutral prompts, and struggle to capture subtle textual differences between similar vulnerable code and patched code.

\section{Enhancement Framework \app{}}
\begin{figure}[htb]
	\centering
	\includegraphics[width=0.6\linewidth]{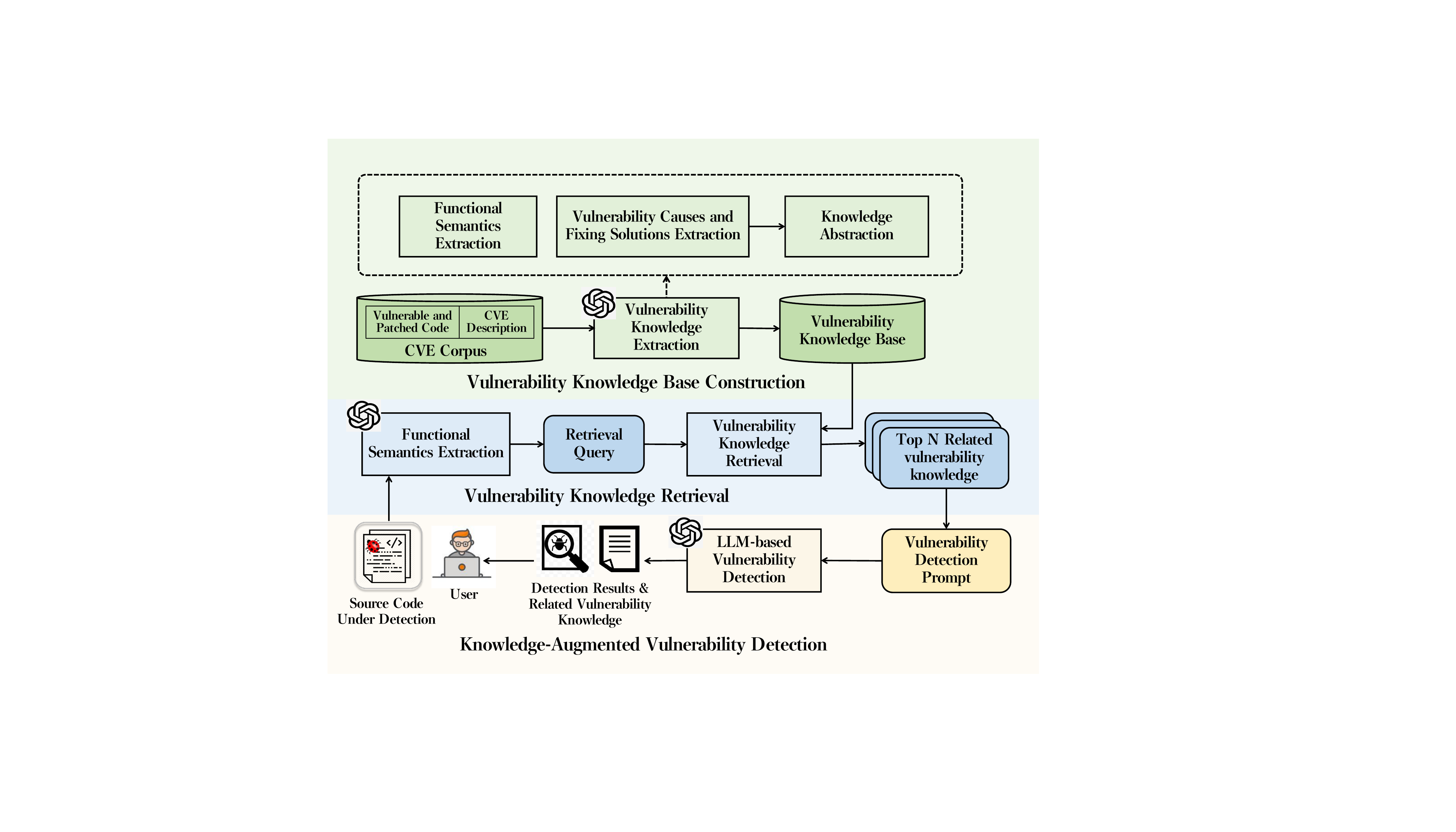}
	\caption{Overview of \app{}}
	\label{fig:overview}
 
\end{figure}

The findings suggest that LLMs require semantic-level guidance for vulnerability detection to avoid relying on superficial code features. Inspired by this, we propose leveraging \textit{high-level vulnerability knowledge} to enhance LLMs in vulnerability detection. Particularly, we propose a novel \textit{knowledge-level} Retrieval-Augmented Generation (RAG) framework \app{} for vulnerability detection, which first distills multi-dimension vulnerability knowledge from existing CVEs and then leverages relevant knowledge items to guide LLM in comprehending the vulnerable behaviors of the given code. As illustrated in Figure~\ref{fig:overview}, \app{} includes three phases: offline vulnerability knowledge base construction, vulnerability knowledge retrieval, and knowledge-augmented vulnerability detection. 

\vspace{-3mm}

\subsection{Vulnerability Knowledge Base Construction}\label{sec:construct}
\app{} constructs a vulnerability knowledge base by automatically extracting multi-dimension knowledge via LLMs from existing vulnerabilities and fixes. Section~\ref{sec:definition} introduces our novel multi-dimension representation for vulnerability knowledge;  Section~\ref{sec:knowledge_extraction} introduces the automatic pipeline of knowledge extraction.

\begin{figure*}[htb]
	\centering
	\includegraphics[width=0.9\linewidth]{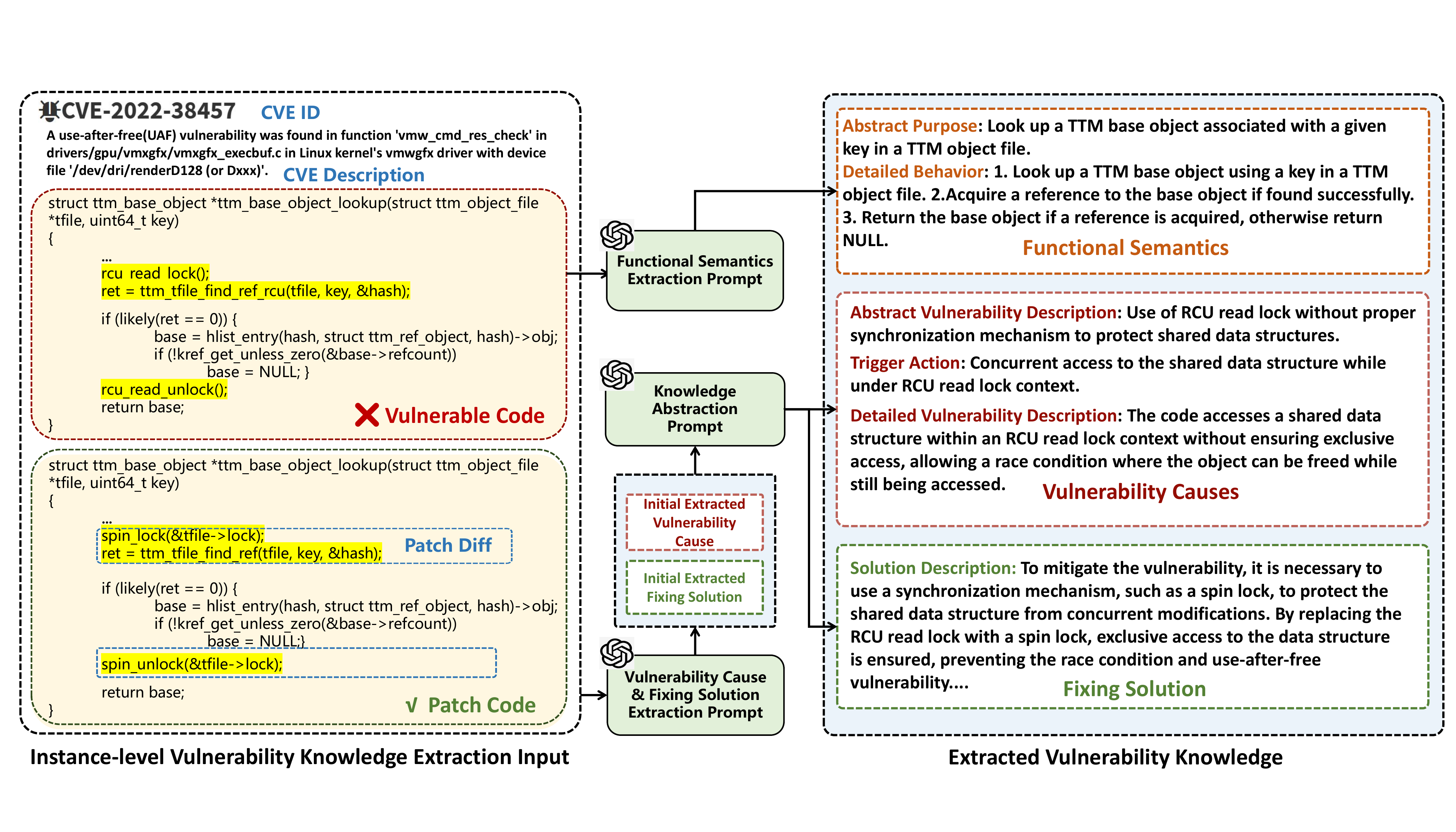}
	\caption{An Example of Vulnerability Knowledge Extraction from CVE-2022-38457}
	\label{fig:example}
        
\end{figure*}
\subsubsection{Vulnerability Knowledge Representation}\label{sec:definition}
Inspired by how developers understand vulnerabilities, we propose a multi-dimensional representation, including seven elements from three dimensions, to describe each vulnerability as follows.

\begin{itemize}[leftmargin=2pt, topsep=2pt, itemsep=0pt]
\item \textbf{Functional Semantics.} This dimension summarizes the high-level functionality (\ie what this code is doing) of the vulnerable code: (1) \textit{Abstract purpose} is the brief summary of the code intention; and (2) \textit{Detailed behavior} is the detailed description of the code behavior.

\item \textbf{Vulnerability Causes.} It describes the reasons for triggering vulnerable behaviors by comparing the vulnerable code and its corresponding patch. The cause can be described from three perspectives: (1)\textit{Triggering action} describes the direct action triggering the vulnerability; (2) \textit{Abstract vulnerability description} is  the brief summary of the cause; and (3) \textit{Detailed vulnerability description} is more concrete descriptions of the causes. 

\item \textbf{Fixing Solutions.} It summarizes the fixing of the vulnerability by comparing the vulnerable code and its corresponding patch.
\end{itemize}

\textit{Functional semantics} are summarized from the vulnerable code, which describe code contexts where vulnerability occurs and are used to facilitate the subsequent retrieval process (Section~\ref{sec:retrieve}); \textit{vulnerability causes} and \textit{fixing solutions} are summarized from the pair of vulnerable  and patched code, which are used to facilitate the subsequent online detection process (Section~\ref{sec:detection}). 
Figure~\ref{fig:example} exemplifies the multi-dimension representation for the real-world vulnerability  CVE-2022-38457.

\subsubsection{Knowledge Extraction} \label{sec:knowledge_extraction}
For each existing CVE instance (including a pair of vulnerable and patched code and its CVE description), \app{} first leverages LLM to extract each dimension of knowledge; then \app{} performs a knowledge abstraction step to increase the generality of extracted knowledge items. 



\parabf{Functional Semantics Extraction.}
Given the vulnerable code, \app{} prompts LLMs to extract both its abstract purpose and  detailed behavior. The detailed prompt is in Appendix~\ref{ap:prompt1}.






\parabf{Vulnerability Causes and Fixing Solutions Extraction.} 
As the causes and fixing solutions are often logically connected, \app{} extracts two dimensions together to maximize the reasoning capabilities of  LLMs.
Given a pair of vulnerable and patched code, \app{} incorporates two rounds to extract the vulnerability causes and the corresponding fixing solutions. In the first round, \app{} instructs LLMs to explain the modification from vulnerable code to patched code; in the second round, \app{} further asks LLMs to extract relevant information in dimensions of causes and fixing solutions based on the explanations generated in the first round. Such a two-step strategy follows a CoT paradigm, which inspires LLM reasoning capabilities by thinking step-by-step~\cite{wei2023chainofthought, zhang2022automatic, li2023structured, DBLP:journals/corr/abs-2402-17230}. Additionally, \app{} includes two shots of demonstration examples to guide the output formats of LLMs. The detailed prompts for vulnerability causes and fixing solutions extraction are in Appendix~\ref{ap:prompt1}.

\parabf{Knowledge Abstraction.}
As different vulnerability instances might share high-level commonality (\eg the similar causes and fixing solutions), \app{} further performs abstraction to distill more general knowledge representation that is less bonded to concrete code implementation details. Particularly, \app{} leverages LLMs to abstract the concrete code elements (\ie method invocations, variable names, and types) in the extracted vulnerability causes and fixing solutions. 
Detailed prompts for knowledge abstraction are in Appendix~\ref{ap:prompt1}. We further illustrate two abstraction guidelines as follows.
\begin{itemize}[left=0pt, topsep=1pt, itemsep=1pt, parsep=0pt]
\item  \textit{Abstracting Method Invocations.} The extracted knowledge might contain concrete method invocations with detailed function identifiers (\eg \texttt{io\_worker\_handle\_work function}) and parameters (\eg \texttt{mutex\_lock(\&dmxdev->mutex)}), which can be abstracted into the generalized description (\eg ``during handling of IO work processes'' and ``employing a locking mechanism akin to mutex\_lock()'').  

\item \textit{Abstracting Variable Names and Types.} The extracted knowledge might contain concrete variable names or types (\eg  
``without \&dev->ref initialization''), which can be abstracted into the more general description (\eg ``without proper reference counter initialization'').
\end{itemize}

\parabf{Vulnerability Knowledge Base.}
For each vulnerability instance, \app{} generates a multi-dimensional knowledge item  with the knowledge extraction and abstraction described above. All the knowledge items are aggregated to form the final \textit{vulnerability knowledge base}. 
In our experiments, to construct the vulnerability knowledge base, we use the remaining 1,462 pairs of vulnerable and patched code that are not selected into our benchmark \ourbench{} (Section~\ref{sec:benchmark}), ensuring that there is no data overlap between the evaluation benchmark and the knowledge base. Detailed statistics within the knowledge base are in Appendix~\ref{ap:dataset}.

\subsection{ Vulnerability Knowledge Retrieval}\label{sec:retrieve}
For a given code snippet under detection, \app{} retrieves relevant vulnerability knowledge items from the constructed vulnerability knowledge base in a three-step retrieval process: semantic query generation, candidate knowledge retrieval, and candidate knowledge re-ranking.

\parabf{Semantic Query Generation.} Different from existing RAG pipelines for code-related tasks~\cite{wang2024coderagbenchretrievalaugmentcode} that solely use code as the retrieval query, \app{} uses a mixed query of both code and its functional semantics to find the knowledge item that share high-level  functional similarity as the given code.  
\app{} prompts LLMs to extract the functional semantics of the given code, using the method described in Section~\ref{sec:knowledge_extraction}. The abstract purpose, detailed behavior, and  code itself, form the query for the subsequent retrieval. 

\parabf{Candidate Knowledge Retrieval.} \app{} conducts similarity-based retrieval using above three query elements: the code, abstract purpose, and detailed behavior.
It retrieves the Top-n knowledge items (where 
n=10 in our experiments) for each element, resulting in a total of 30 candidate items. Duplicates across query elements are removed to ensure uniqueness.
The retrieval is based on the similarity between each query element and the corresponding elements of the knowledge items. \app{} adopts BM25~\cite{bm25} for similarity calculation, a method widely used in search engines due to its efficiency and effectiveness~\cite{sun2023gpt}.
Before calculating BM25 similarity, both the query and the retrieval documentation undergo standard preprocessing procedures, including tokenization, lemmatization, and stop word removal~\cite{DBLP:conf/iconip/CagatayliC15}.

\parabf{Candidate Knowledge Re-ranking.} 
We re-rank candidate knowledge items with the Reciprocal Rank Fusion (RRF) strategy. For each retrieved knowledge item $k$, we aggregate the reciprocal of its rank across all three query elements. If a knowledge item $k$ is not retrieved by a particular query element, we assign its rank as infinity. 
Detailed formulas and implementation of the retrieval and re-ranking process are in Appendix~\ref{ap:retrieval}. In the end, we keep  Top-10 candidate knowledge items with the highest re-rank scores as the final knowledge items for the subsequent vulnerability detection.

\subsection{Knowledge-Augmented Vulnerability Detection}\label{sec:detection}
Based on the retrieved knowledge items, \app{} leverages LLMs to reason whether the given code is vulnerable. 
However, directly incorporating all the retrieved knowledge items into one prompt can hinder the effectiveness of the models, as LLMs often perform poorly on long  contexts~\cite{DBLP:journals/corr/abs-2307-03172}. Therefore, \app{} iteratively enhances LLMs with each retrieved knowledge item by sequentially checking whether the given code exhibits the same vulnerability cause without the corresponding fixing solutions.

If the given code exhibits the same vulnerability cause as the knowledge item but without applying the relevant fixing solution, it is identified as \textit{vulnerable}. Otherwise, \app{} cannot identify the code as vulnerable with the current knowledge item and proceeds to the next iteration (\ie using the next retrieved knowledge item). If the code cannot be identified as vulnerable with any of the retrieved knowledge items, it is identified as \textit{non-vulnerable}. The iteration process terminates when (1) the code is identified as vulnerable or (2) all the retrieved knowledge items have been considered. 
The detailed prompts of this phase are in Appendix \ref{ap:prompt2}.

\section{Evaluation for \app{}}

\label{sec:eval}
We answer the following RQs to extensively evaluate the effectiveness and usability of \app{}.

\begin{itemize}[left=0pt, topsep=1pt, itemsep=2pt,parsep=2pt]

\item \textbf{RQ3 (Overall Improvements):} How does \app{} improve LLMs in vulnerability detection?

\item \textbf{RQ4 (User Study on Usability):} How is the quality of \app{} generated knowledge? How can the \app generated knowledge help manual vulnerability comprehension? 

\item \textbf{RQ5 (Case Study on Detecting New Vulnerabilities):} Can the \app{} generated knowledge help detect previously-unknown vulnerabilities in real-world software systems?
\end{itemize}

\paragraph{Implementation.} During the offline knowledge base construction, we employ GPT-3.5-turbo-0125~\cite{chatgpt}, given its rapid response and cost-effectiveness in generating a large volume of vulnerability-related knowledge items~\cite{sun2023gpt}. For the online knowledge retrieval, we use Elasticsearch~\cite{elasticsearch} as our search engine. We use 2317 vulnerable-patched pairs across 1154 CVEs that do not overlap with \ourbench{} as the training set for knowledge base Construction. Detailed dataset division and training data statistics are in Appendix \ref{ap:dataset}.
For the online knowledge-augmented detection, we study the same four LLMs (GPT-4o, Claude Sonnet 3.5, Qwen2.5-Coder-32B-Instruct, and DeepSeek-R1) as in the study. \app{} and all studied baselines are trained and tested on a Linux server (CPU: Intel Xeon Platinum 8358P, GPU: NVIDIA A800 Tensor Core GPU).

\subsection{RQ3: Overall Improvements}\label{sec:rq3}
\parabf{Baselines.} Besides the basic prompt and three advanced prompts studied in RQ1 and RQ2, 
we further include a \textbf{code-level RAG} baseline and a \textbf{fine-tuning} based baselines. 

\textbf{Code-based RAG} enhances basic LLMs by retrieving similar code snippets from training datasets to enrich the prompts. In particular, comparing \app{} with code-level RAG can investigate the contribution of our knowledge-level representation. The detailed prompt design of code-level RAG is in Figure \ref{fig:mov_example} (b) in Appendix \ref{ap:case_rq2}. Comparing \app{} with code-level RAG can investigate the contribution of our knowledge-level representation. 

\textbf{LLMAO}~\cite{DBLP:conf/icse/YangGMH24} fine-tunes the LLM on the Devign dataset~\cite{DBLP:conf/nips/ZhouLSD019} for vulnerability detection. 
We directly utilize the public implementation of this baseline. To adapt the techniques to our
benchmark, we (i) replace the original codegen-16b model with the more advanced Qwen2.5-Coder-32b and DeepSeek-R1-7b model and (ii) fine-tune the baseline on our training set for 10 epochs.
Additionally, as LLMAO is a line-level vulnerability detection technique, we adapt it to the function level by regarding the function that has any line with higher-than-0.5 suspiciousness scores as vulnerable.

\begin{table}[htb]
    \centering
    \caption{Effectiveness of \app{}} \label{table:rq3}
    \footnotesize
    \begin{adjustbox}{width=0.55\linewidth}
        \begin{tabular}{c|c|c|c|c}
        \hline
        \textbf{Tech.} & \textbf{LLM} & \textbf{Pair Acc.} & \textbf{Bal. Recall} & \textbf{Bal. Pre.}  \\ \hline

        \multirow{4}{*}{Code RAG
        }& GPT-4o & 0.05  & 0.51 & 0.54 \\
        & Claude & 0.11  & 0.52 & 0.53  \\
        & Qwen & 0.09  & 0.54 & 0.59 \\
        & DeepSeek & 0.05  & 0.52 & 0.60\\
        \hline

        \multirow{2}{*}{LLMAO}
        & Qwen & 0.04 & 0.51 & 0.51  \\
        & DeepSeek & 0.04 & 0.51 & 0.51 \\ \hline
        
        \multirow{4}{*}{\app{}
        }& GPT-4o & \textbf{0.32} & 0.58 & \textbf{0.63}\\
        & Claude & 0.27  & \textbf{0.61} & 0.62 \\
        & Qwen & 0.26  & 0.59 & 0.61 \\
        & DeepSeek & 0.30 & \textbf{0.61} & 0.62 \\ \hline  
     
        \end{tabular}
    \end{adjustbox}
\end{table}

\parabf{Results.}
Table~\ref{table:rq3} compares \app{} with code-level RAG and LLMAO on \ourbench{}. Due to space limits, here we do not repeat the results of other baselines (in Table~\ref{table:rq1} and Table~\ref{table:rq2}). 
Detailed comparison between \app{} and baselines in each CWE category is in Appendix~\ref{ap:rq3} and the bad case analysis of \app{} is in Appendix~\ref{ap:bad_case}.
Overall, \app{} substantially outperforms all baselines in all metrics. Particularly, \app{} not only improves the pair accuracy of LLMs (with 16\% - 24\% increase) but also improves the balanced precision and recall by 9\%-14\% and 7\%-11\%. 

Compared to code-level RAG, \app{} shows greater effectiveness in enhancing LLMs for vulnerability detection, with consistent improvements across all metrics. This highlights the contribution of our novel vulnerability knowledge representation and underscores the superiority of knowledge-level RAG over code-level RAG. We manually inspect cases where \app{} successfully identifies vulnerable and patched code pairs that code-level RAG fails. We identify two key reasons for the superior performance of \app{}. 
(1) In the retrieval phase, knowledge-level RAG more accurately retrieves semantically relevant vulnerabilities from the knowledge base, whereas code-level RAG often retrieves textually similar but semantically irrelevant vulnerabilities. As a result, the vulnerabilities retrieved by code-level RAG offer limited utility for or even mislead LLMs in vulnerability detection.
(2) In the inference phase, even when retrieving the same vulnerabilities, the high-level representation of vulnerability knowledge provided by \app{} can more accurately prompt LLMs while the plain representation of code pairs used in code-level RAG cannot.  
Appendix~\ref{ap:case_rq2} presents such two cases observed in our experiments.

\subsection{RQ4: Usability for Developers}
\label{sec:user_study}
We conduct a user study to investigate the quality of \app{} generated knowledge and whether the knowledge can help developers understand and check the vulnerabilities in code. 

\parabf{Tasks and Participants.} 
We select 10 cases (5 true and 5 false positives) from \ourbench{} for the user study. We invite 6 participants with 3-5 years c/c++ programming experience.  Participants are tasked to identify whether the given code is vulnerable in two settings. (1) Basic setting: provided with the code and the detection labels generated by \app{}; (2) Knowledge-accompanied setting: provided with the basic setting and \app{} generated vulnerability knowledge.
Detailed procedure and scoring criteria are in Appendix~\ref{ap:rq4}.

\parabf{Metrics.} 
Beyond recording user outputs (\ie{} vulnerable or not) of each case, we further survey the participants on the helpfulness, preciseness, and generalizability of the vulnerability knowledge on a 4-point Likert scale~\cite{Likert1932} (\ie{} 1-disagree; 2-somewhat disagree; 3-somewhat agree; 4-agree).

\parabf{Results.}
Participants rate the helpfulness, preciseness, and generalizability with average scores of 3.00, 3.20,  and 2.97, respectively. It indicates the high quality of vulnerability knowledge generated by \app{}. 
Additionally, participants provided with \app{} generated vulnerability knowledge can more precisely identify the vulnerable and non-vulnerable code with a statistically significant improvement (\ie{} 77\% detection accuracy with knowledge v.s. 60\% detection accuracy without knowledge, $p$ = 0.01). It confirms the usability of \app{} generated knowledge for manual vulnerability comprehension.

\subsection{RQ5: Detecting New Vulnerabilities}
We investigate whether \app{} generated vulnerability knowledge can detect previously-unknown vulnerabilities in real-world software systems. In particular, we apply \app{} on the recent Linux Kernel release (v6.9.6, June 2024) given the importance of Kernel systems. Given the large scale of Linux kernels, we randomly sample a set of files within the \textit{drivers} component, including 1,568 functions in total. We apply \app{} with GPT-4 on the 1,568 functions. \app{} detects \detectbugnum{} previously-unknown bugs, and \confirmbugnum{} of them have been confirmed as real bugs by the Linux community with
assigned CVEs. 

Moreover, since \app{} not only generates the detection labels (\ie{} vulnerable or not) but also provides vulnerability knowledge with relevant vulnerability causes and fix suggestions, it is helpful for us writing high-quality bug-reporting emails. For the \confirmbugnum{} confirmed bugs, we further submitted patches based on the fix solutions provided by \app{}, all of which have already been accepted. 
Appendix~\ref{ap:unknown_bug} presents an example of our confirmed bug.

\vspace{-1mm}
\section{Conclusion}
This work reveals the limitation of LLMs in distinguishing between vulnerable and patched code, and proposes a novel knowledge-level RAG framework \app{}, which enhances LLMs with multi-dimensional vulnerability
knowledge distilled from historical vulnerabilities and fixes. \app{} outperforms all baselines in vulnerability detection; and \app{}  generated knowledge improves manual vulnerability detection by 17\% accuracy increase. Additionally, \app{} detects \detectbugnum{} previously-unknown bugs in the Linux kernel and 6 of them have been confirmed by the Linux community with assigned
CVEs.

\bibliography{ref}
\bibliographystyle{abbrv}

\newpage
\appendix

\section{Dataset }
\label{ap:dataset}

\parabf{Construction Procedure.} In this section, we elaborate more details on the following two key steps involved in the benchmark construction process.

\textit{Vulnerable and Patched Code Collection.} 
We first collect all the CVEs from ~\cite{linuxcve}, an open-source project dedicated to automatically tracking CVEs within the upstream Linux kernel. Based on the list of collected CVE IDs, we further extract corresponding CWE IDs and CVE descriptions from the National Vulnerability Database (NVD), enriching our dataset with detailed vulnerability categorizations and descriptions.
Based on the CVE ID list, we then parse the commit information for each CVE to extract function-level vulnerable and patched code pairs. Vulnerable code snippets prior to the commit diffs are labeled as positive samples and the patched code snippets as negative samples. 
In this way, we initially obtain a dataset of 4,667 function pairs of vulnerable and patched code across 2,174 CVEs.

\textit{Patched Code Verification.} The patched code cannot always be non-vulnerable, thus it is important to double-check the correctness of the patched code. To this end, we further implement a filtering process to ensure the patched code is not subsequently reverted or modified by other commits. 
Specifically, we construct a patch graph in which vulnerable and patched code pairs are represented as independent triplets. Each triplet consists of a head node representing the vulnerable code and a tail node representing its corresponding patched code. If the patched code is modified or reverted by subsequent commits, the triplet evolves into a chain or a loop comprising multiple nodes. For chains in the graph, we retain only the vulnerable and patched code pairs linked by the final edge. For loops, we eliminate all nodes within the loop. This process systematically filters out all patched code snippets that have been altered or reverted, ensuring the correctness of our benchmark dataset.

\parabf{Data Statistics.}
We focus on the Top-10 most prevalent CWEs to construct our datasets.
The detailed information for each CWE is presented in Table \ref{table:cwe_def}.
Consequently, we obtain 2903 function pairs of vulnerable-patched pairs across 1325 CVEs. We then randomly divide these data into 1:4 for the test set and training set, with the test set serving as our benchmark dataset \ourbench{}. PairVul includes 586 pairs across 420 CVEs.   The statistical data for each CWE category within \ourbench{} are detailed in Table \ref{table:benchmark}. The training set is used to construct our knowledge base, including 2317 pairs of vulnerable and patched code across 1154 CVEs  that do not overlap with \ourbench{}. Table \ref{table:training_data} presents the distribution of the five CWE categories within the knowledge base.

\begin{table}[htb]
	\centering
    \caption{CWE Definition}
    \label{table:cwe_def}
    \footnotesize
    \begin{adjustbox}{width=1.0\linewidth}
    \begin{tabular}{m{2cm}<{\centering}|m{5cm}<{\centering}|m{8cm}<{\raggedright}} \hline
    \textbf{CWE} & \textbf{Name} & \textbf{Definition} \\  \hline
    CWE-416 & Use After Free & The product reuses or references memory after it has been freed.\\ \hline 
    CWE-476 & NULL Pointer Dereference & The product dereferences a pointer that it expects to be valid but is NULL. \\ \hline
    CWE-362 & Race Condition & Concurrent code sequences can simultaneously access a shared resource without proper synchronization.\\ \hline
    CWE-119 & Improper Restriction of Operations within the Bounds of a Memory Buffer & The program accesses memory outside its allocated buffer boundaries.\\ \hline
    CWE-787 & Out-of-bounds Write & The product writes data past the end, or before the beginning, of the intended buffer. \\ \hline
    CWE-20 & Improper Input Validation & Input validation is missing or inadequate for safe data processing.\\ \hline
    CWE-200 & Exposure of Sensitive Information to an Unauthorized Actor & The product leaks sensitive data to unauthorized parties.\\ \hline
    CWE-125 & Out-of-bounds Read & The product reads data past the end, or before the beginning, of the intended buffer.\\ \hline
    CWE-264 & Permissions, Privileges, and Access Controls & Access control issues arise from poor management of permissions, privileges, and security features. \\ \hline
    CWE-401 & Missing Release of Memory after Effective Lifetime & The product has memory leaks due to poor deallocation practices.\\ \hline

    \end{tabular}
    \end{adjustbox}
\end{table}

\begin{table}[htp]
    \footnotesize
    \centering
    \caption{Statistics of \ourbench{}}
    \label{table:benchmark}
    \begin{adjustbox}{width=1.0\linewidth}
    \begin{tabular}{c|c|c|c|c|c|c|c|c|c|c}
        \hline
        \textbf{CWE} & \textbf{CWE-416} & \textbf{CWE-476} & \textbf{CWE-362} & \textbf{CWE-119} & \textbf{CWE-787} &\textbf{CWE-20} &\textbf{CWE-200} &\textbf{CWE-125}&\textbf{CWE-264} &\textbf{CWE-401} \\
        \hline
        \textbf{CVE Num.} & 117 & 58 & 53 & 36 & 40 & 36 &31 &29 &13 &23\\
        \textbf{Pair Num.} & 166 & 71 & 81 & 44 & 47 & 46 &39 &35 &31 &26\\
        \hline
    \end{tabular}
    \end{adjustbox}
\end{table}

\begin{table}[htp]
    \footnotesize
    \centering
    \caption{Statistics of Training Set}
    \label{table:training_data}
    \begin{adjustbox}{width=1.0\linewidth}
    \begin{tabular}{c|c|c|c|c|c|c|c|c|c|c}
        \hline
        \textbf{} & \textbf{CWE-416} & \textbf{CWE-476} & \textbf{CWE-362} & \textbf{CWE-119} & \textbf{CWE-787} &\textbf{CWE-20}&\textbf{CWE-200}&\textbf{CWE-125}&\textbf{CWE-264}&\textbf{CWE-401}\\
        \hline
        \textbf{CVE Num.} & 300 & 163 & 159 & 111 & 107 & 79 & 92  & 89 & 41 & 76\\
        \textbf{Pair Num.} & 660 & 281 & 320 & 173 & 187 & 182 &152 &140 & 120 & 101\\
        \hline
    \end{tabular}
    \end{adjustbox}
\end{table}

\parabf{Data Format.} Our benchmark \ourbench{} contains the following information for each vulnerability. (i) \textit{}{CVE ID}: the unique identifier assigned to a reported vulnerability in the Common Vulnerabilities and Exposures (CVE); (ii) \textit{CVE Description}: descriptions of the vulnerability provided by the CVE system, including the manifestation, the potential impact, and the environment where the vulnerability may occur; (iii) \textit{CWE ID}: the Common Weakness Enumeration identifier that categorizes the type of the vulnerability exploits; (iv) \textit{Vulnerable Code}: the source code snippet containing the vulnerability, which will be modified in the commit; (v) \textit{Patched Code}: the source code snippet that has been committed to fix the vulnerability in the vulnerable code; (vi) \textit{Patch Diff}: a detailed line-level difference between the vulnerable and patched code with added and deleted lines.

\section{Studied Baselines}
\label{ap:baseline}

In RQ1, we evaluate the capabilities of studied LLMs with the following basic prompt:

\begin{mdframed}
[linecolor=myblue!50,linewidth=2pt,roundcorner=10pt,backgroundcolor=myyellow!20]
\small
\hspace{3.5mm}\textbf{Basic Prompt:}
Is this code vulnerable? Answer in Yes or No.

\#\#\# Code Snippet: 
\textit{[Code Snippet]}.

\end{mdframed}

In RQ2, we further investigate three state-of-the-art
prompting strategies, including:

\textbf{Chain-of-thought(CoT) strategies} enhances the basic LLMs with two chain-of-thought prompt design, guiding LLMs step-by-step reasoning.

\begin{mdframed}
[linecolor=myblue!50,linewidth=2pt,roundcorner=10pt,backgroundcolor=myyellow!20]
\small
\hspace{3.5mm}\textbf{CoT-1 Prompt:}  
I want you to act as a vulnerability detection expert. Initially, you need to explain the behavior of the code. Subsequently, you need to determine whether the code is vulnerable. Answer in YES or NO.

\#\#\# Code Snippet: 
\textit{[Code Snippet]}.

\textbf{CoT-2 Prompt:} I want you to act as a vulnerability detection system. Initially, you need to explain the behavior of the given code. Subsequently, analyze whether there are potential root causes that could result in vulnerabilities. Based on above analysis, determine whether the code is vulnerable, and conclude your answer with either YES or NO.

\#\#\# Code Snippet: 
\textit{[Code Snippet]}.

\end{mdframed}

\textbf{CWE-enhanced strategies} enhances the basic LLM by incorporating CWE description information~\cite{cwe} as vulnerability knowledge to LLMs.

\begin{mdframed}
[linecolor=myblue!50,linewidth=2pt,roundcorner=10pt,backgroundcolor=myyellow!20]
\small
\hspace{3.5mm}\textbf{CWE-enhnced Prompt:} 
I want you to act as a vulnerability detection system. I will provide you with a code snippet and a CWE description. Please analyze the code to determine if it contains the vulnerability described in the CWE. Answer in YES or NO.

\#\#\# Code Snippet: 
\textit{[Code Snippet]}.

\#\#\# CWE Description: \textit{[CWE Description]}
\end{mdframed}

\section{Prompt Design of \app{}}
\label{ap:prompt}

\subsection{Prompt Templates in Vulnerability Knowledge Base Construction}
\label{ap:prompt1}

Given the vulnerable code snippet, \app{} prompts LLMs with the following instructions to summarize both the abstract purpose and the detailed behavior respectively, where the placeholder ``\texttt{[Vulnerable Code]}'' denotes the vulnerable code snippet.

\begin{mdframed}
[linecolor=myblue!50,linewidth=2pt,roundcorner=10pt,backgroundcolor=myyellow!20]
\small
\hspace{3.5mm}\textbf{Prompt for Abstract Purpose Extraction:}  \texttt{[Vulnerable Code]} What is the purpose of the function in the above code snippet? Please summarize the answer in one sentence with the following format: ``Function purpose:''.

\textbf{Prompt for Detailed Behavior Extraction:} \texttt{[Vulnerable Code]} Please summarize the functions of the above code snippet in the list format without any other explanation: ``The functions of the code snippet are: 1. 2. 3...''
\end{mdframed}

The detailed prompts for vulnerability causes and fixing solutions extraction
are as follows, where the placeholders ``\texttt{[Vulnerable Code]}'', ``\texttt{[Patched Code]}'',  and ``\texttt{[Patch Diff]}'' denote the vulnerable code, the patched code, and the code diff of the given vulnerability, and \textit{[CVE ID]} and \textit{[CVE Description]} denote the details of the given vulnerability. 

\begin{mdframed}
[linecolor=myblue!50,linewidth=2pt,roundcorner=10pt,backgroundcolor=myyellow!20]
\small

\hspace{3.5mm}\textbf{Extraction Prompt in Round 1:} This is a code snippet with a vulnerability \textit{[CVE ID]}:
\texttt{[Vulnerable Code]}
The vulnerability is described as follows:\textit{[CVE Description]}
The correct way to fix it is by \textit{[Patch Diff]}
The code after modification is as follows: \textit{[Patched Code]}
Why is the above modification necessary?

\textbf{Extraction Prompt in Round 2:}
I want you to act as a vulnerability detection expert and organize vulnerability knowledge based on the above vulnerability repair information. Please summarize the generalizable specific behavior of the code that leads to the vulnerability and the specific solution to fix it. Format your findings in JSON.
Here are some examples to guide you on the level of detail expected in your extraction:
\textit{[Vulnerability Causes and Fixing Solution Example 1]}
\textit{[Vulnerability Causes and Fixing Solution Example 2]}
\end{mdframed}

The detailed prompts for knowledge abstraction are as follows, which queries LLMs to abstract the method invocations and variable names. 

\begin{mdframed}
[linecolor=myblue!50,linewidth=2pt,roundcorner=10pt,backgroundcolor=myyellow!20]
\small

\hspace{3.5mm}\textbf{Knowledge Abstraction Prompt: }
With the detailed vulnerability knowledge extracted from the previous stage, your task is to abstract and generalize this knowledge to enhance its applicability across different scenarios. Please adhere to the following guidelines and examples provided:

\textit{[Knowledge Abstraction Guidelines and Examples]}
...
\end{mdframed}

\subsection{Prompt Templates in Knowledge-Augmented Vulnerability Detection}
\label{ap:prompt2}

The prompts used for identifying the existence of vulnerability causes and the fixing solutions of the given code snippets are as follows.

\begin{mdframed}
[linecolor=myblue!50,linewidth=2pt,roundcorner=10pt,backgroundcolor=myyellow!20]
\small
\hspace{3.5mm}\textbf{Prompt for Finding Vulnerability Causes:} Given the following code and related vulnerability causes, please detect if there is a vulnerability caused in the code. \textit{[Code Snippet]}. In a similar code scenario, the following vulnerabilities have been found: \textit{[Vulnerability causes][fixing solutions]}. Please use your own knowledge of vulnerabilities and the above vulnerability knowledge to detect whether there is a vulnerability in the code.

\textbf{Prompt for Finding Fixing Solutions:} Given the following code and related vulnerability fixing solutions, please detect if there is a vulnerability in the code. \textit{[Code Snippet]}. In a similar code scenario, the following vulnerabilities have been found: \textit{[Vulnerability causes][fixing solutions]}. Please use your own knowledge of vulnerabilities and the above vulnerability knowledge to detect whether there is a corresponding fixing solution in the code.

\end{mdframed}
\section{Retrieval Implementation}
\label{ap:retrieval}

\app{} adopts BM25~\cite{bm25} for similarity calculation in retrieval process. Given a query $q$ and the documentation $d$ for retrieval, BM25 calculates the similarity score between $q$ and $d$ based on the following Equation~\ref{eq:bm25}, where $f(w_{i}, q)$ is the word $w_{i}$'s term frequency in query $q$, $IDF\left(w_{i}\right)$ is the inverse document frequency of word $w_{i}$. The hyperparameters $k$ and $b$ (where k=1.2 and b=0.75) are used to normalize term frequencies and control the influence of document length. 

\begin{equation}
\label{eq:bm25}
\footnotesize
{Sim}_{BM25}(q, d)=\sum_{i=1}^{n} \operatorname {\frac{IDF\left(w_{i}\right) \times f\left(w_{i}, q\right) \times(k+1)}{f\left(w_{i}, q\right)+k \times\left(1-b+b \times \frac{|q|}{avgdl}\right)}}
\end{equation} 
 
We re-rank candidate knowledge items with the Reciprocal Rank Fusion (RRF) strategy. For each retrieved knowledge item $k$, the re-rank score for $k$ is calculated using the following Equation \ref{eq:rerank}. $E$ denotes the set of all query elements (\ie{} the code, the abstract purpose, and the detailed behavior), $rank_{t}(k)$ denotes the rank of knowledge item $k$ based on query element $t$.

\begin{equation}
\label{eq:rerank}
ReRankScore_{k} = \sum_{t\in E}^{} \frac{1}{rank_{t} (k)} 
\end{equation} 
\section{Overall Performance}
\label{ap:rq3}

\begin{figure*}[!htbp]
    \centering   \subfigure[CWE-416 Pair Accuracy Result]{
        \includegraphics[width=0.45\linewidth]{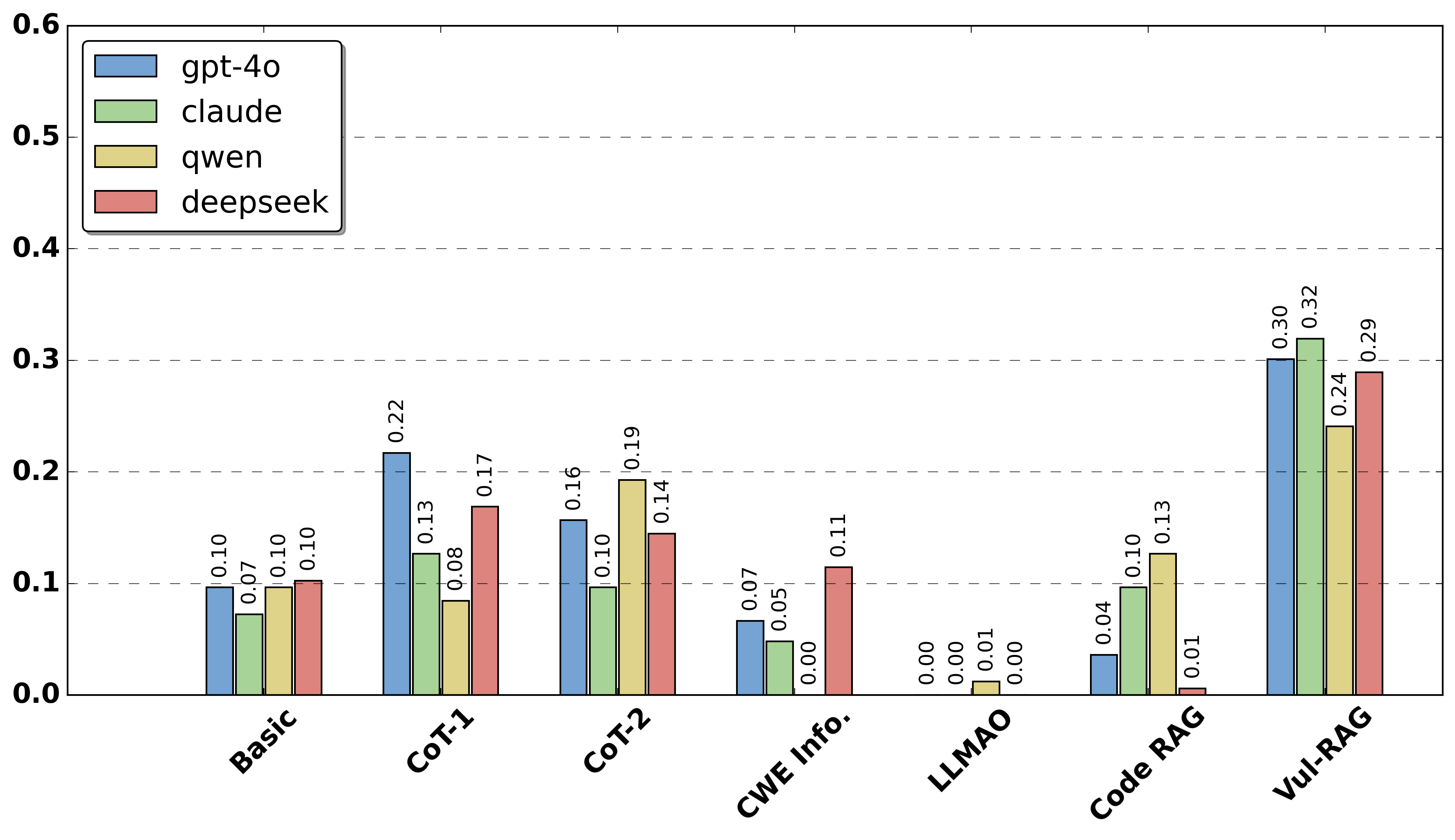}
        \label{fig:416}
    }
    \subfigure[CWE-476 Pair Accuracy Result]{
        \includegraphics[width=0.45\linewidth]{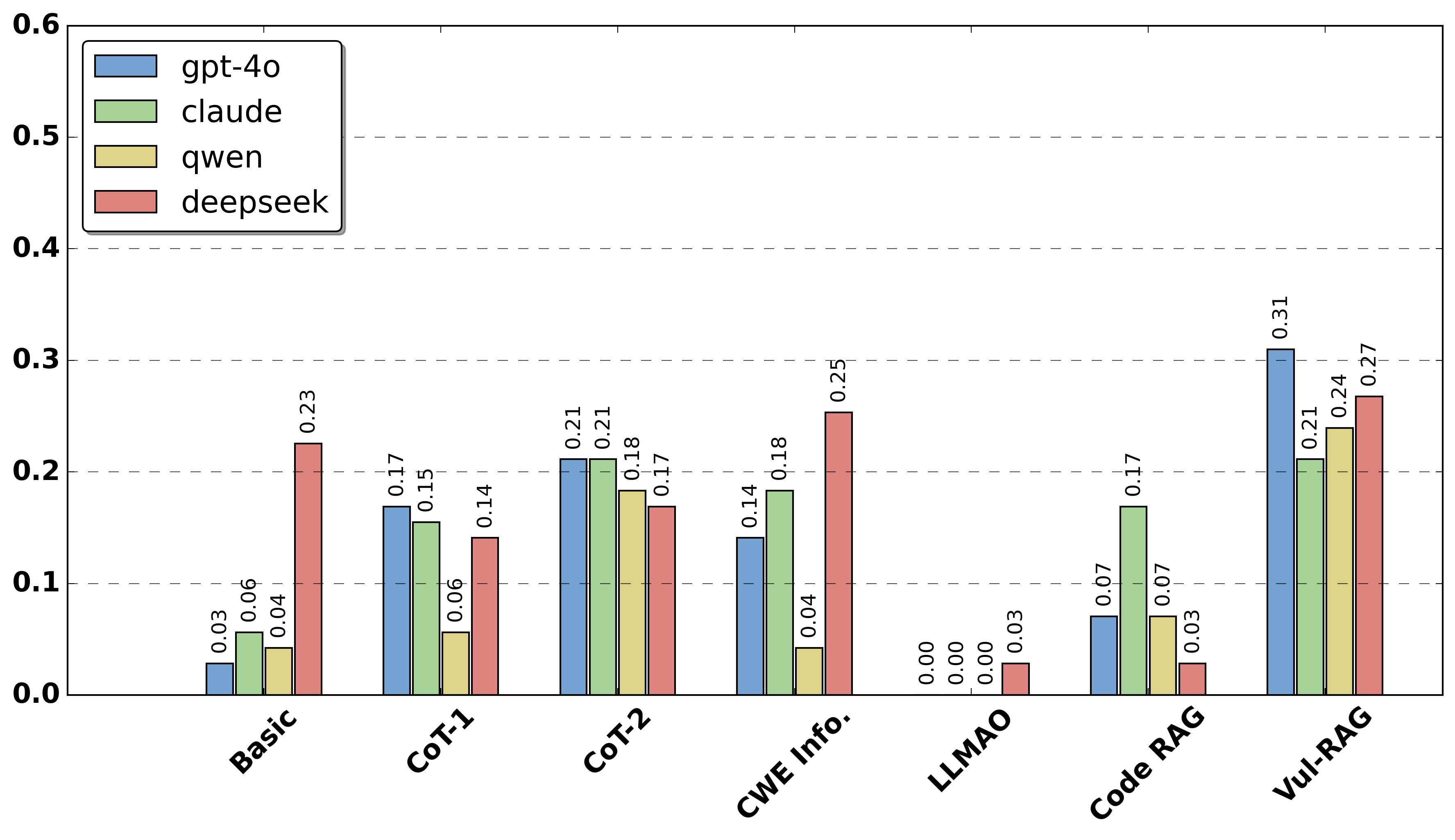}
        \label{fig:476}
    }
        \subfigure[CWE-362 Pair Accuracy Result]{
        \includegraphics[width=0.45\linewidth]{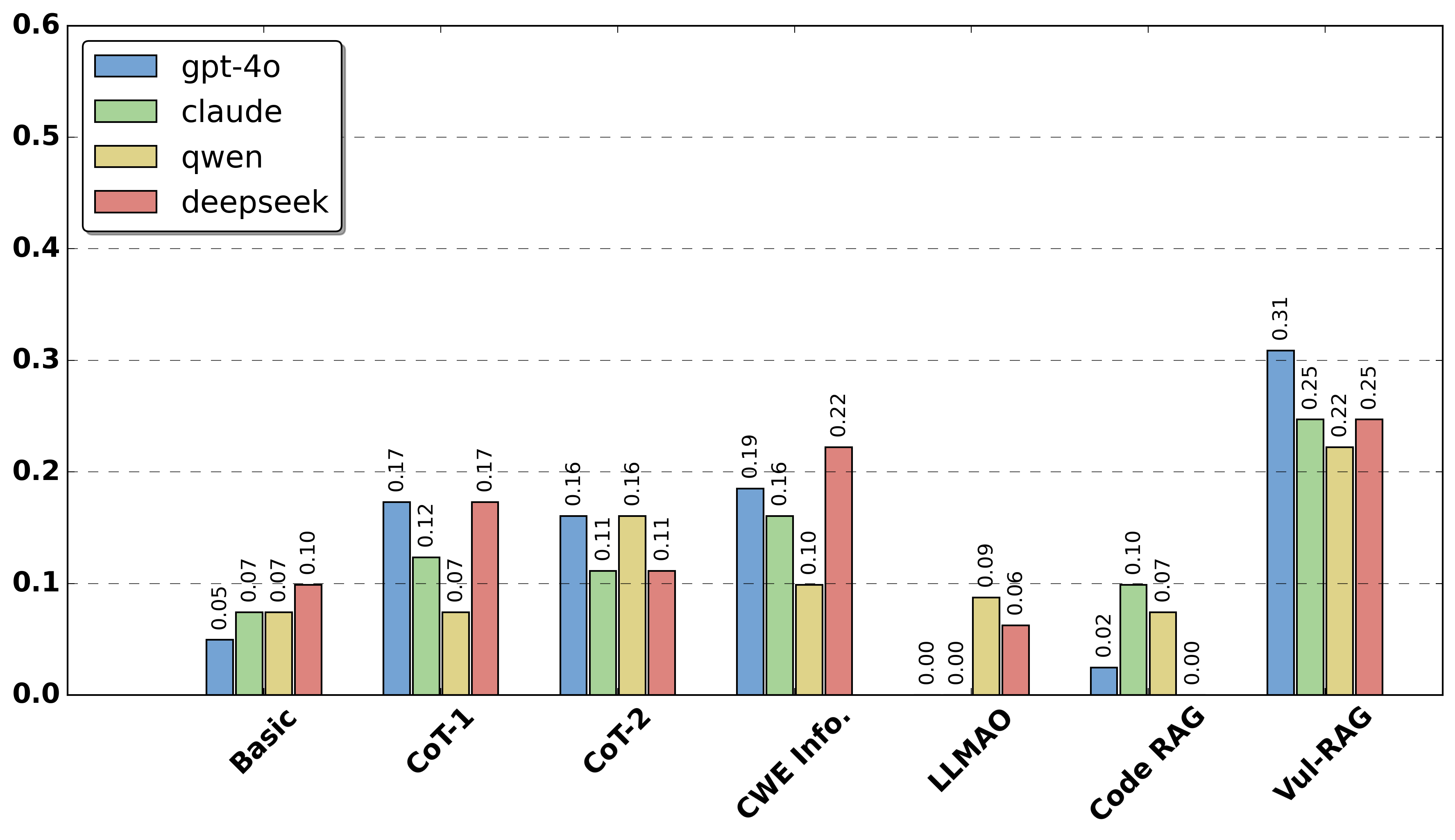}
        \label{fig:362}
    }
    \subfigure[CWE-119 Pair Accuracy Result]{
        \includegraphics[width=0.45\linewidth]{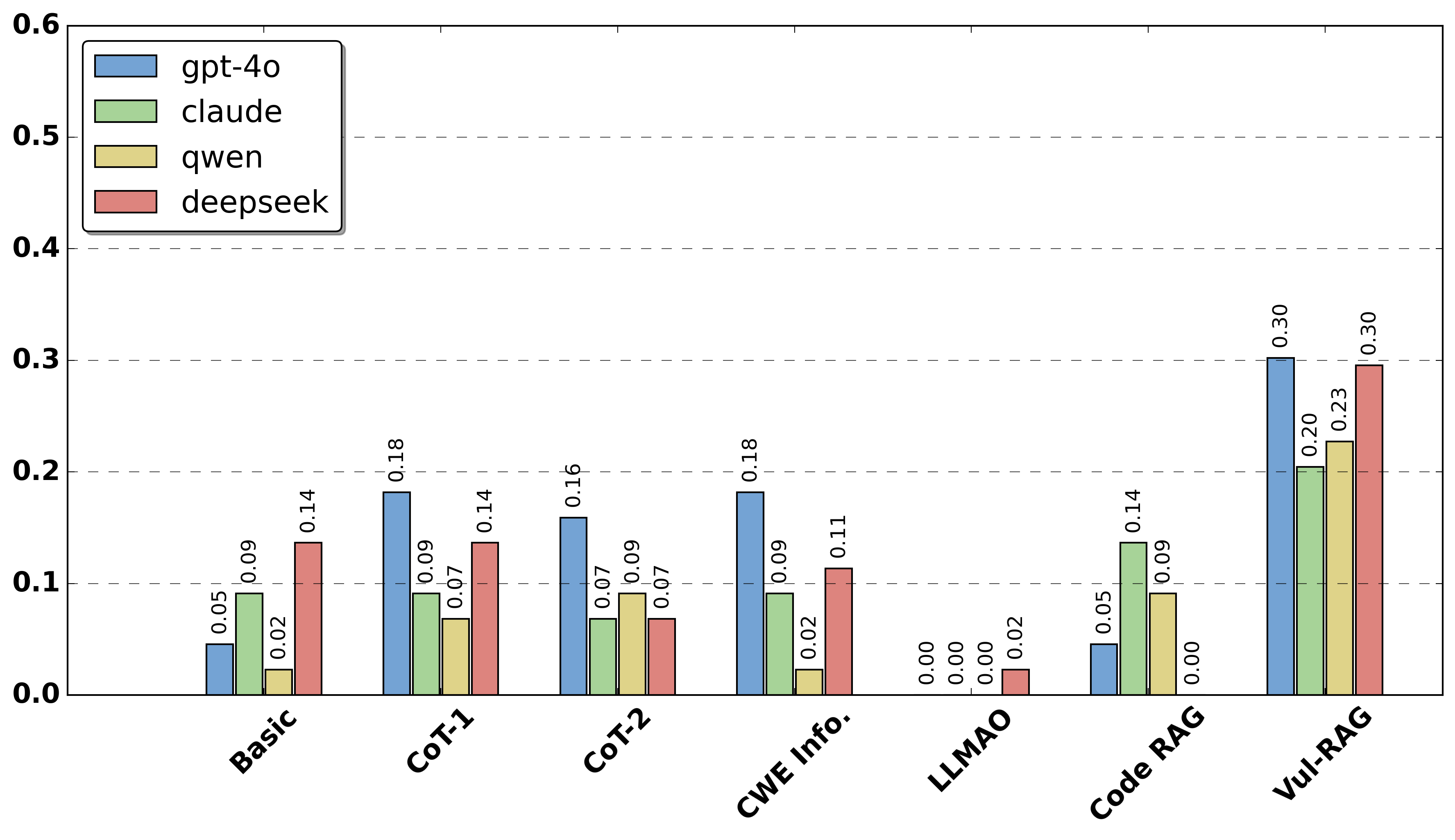}
        \label{fig:119}
    }
        \subfigure[CWE-787 Pair Accuracy Result]{
        \includegraphics[width=0.45\linewidth]{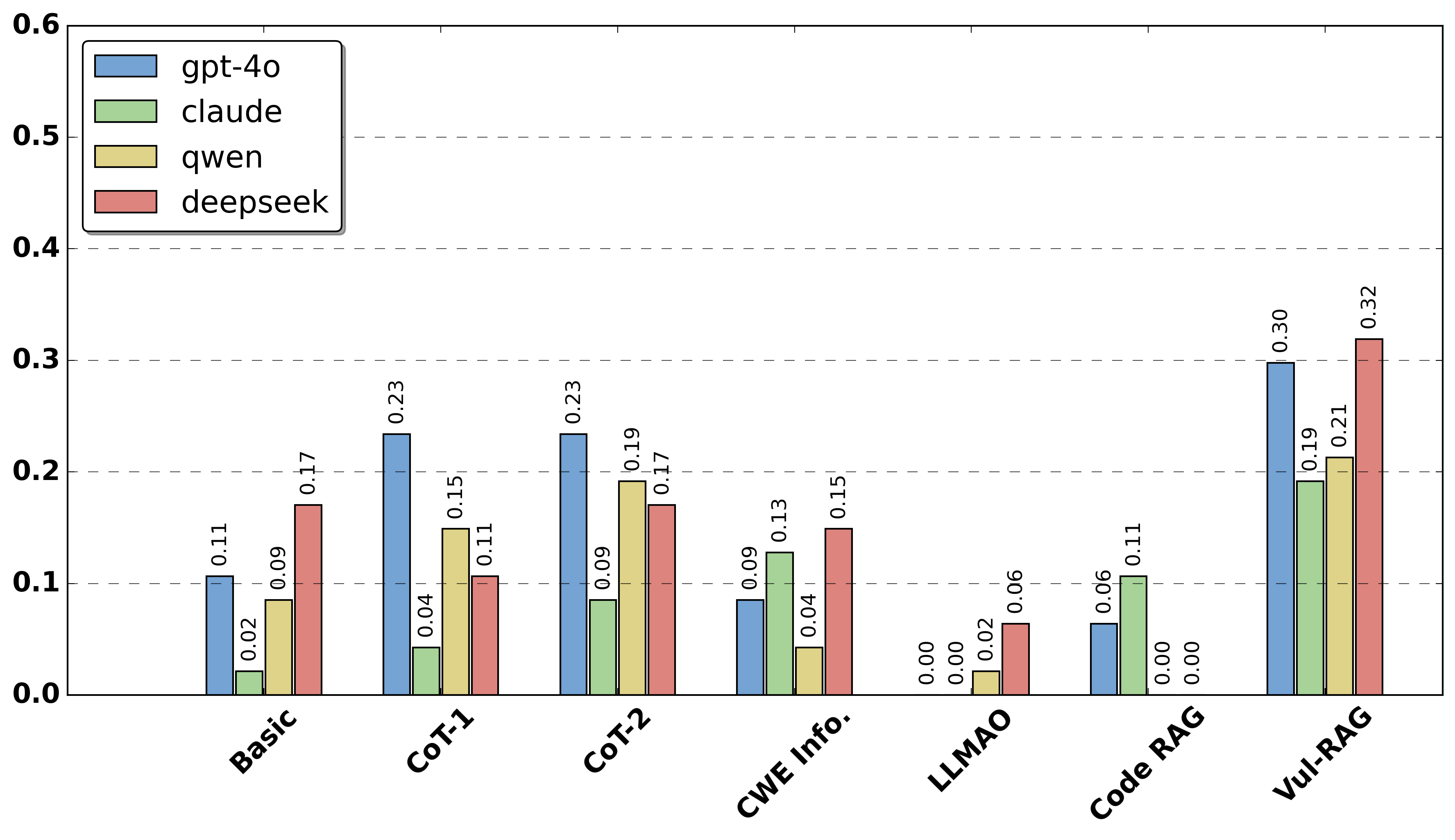}
        \label{fig:787}
    }
    \subfigure[CWE-20 Pair Accuracy Result]{
        \includegraphics[width=0.45\linewidth]{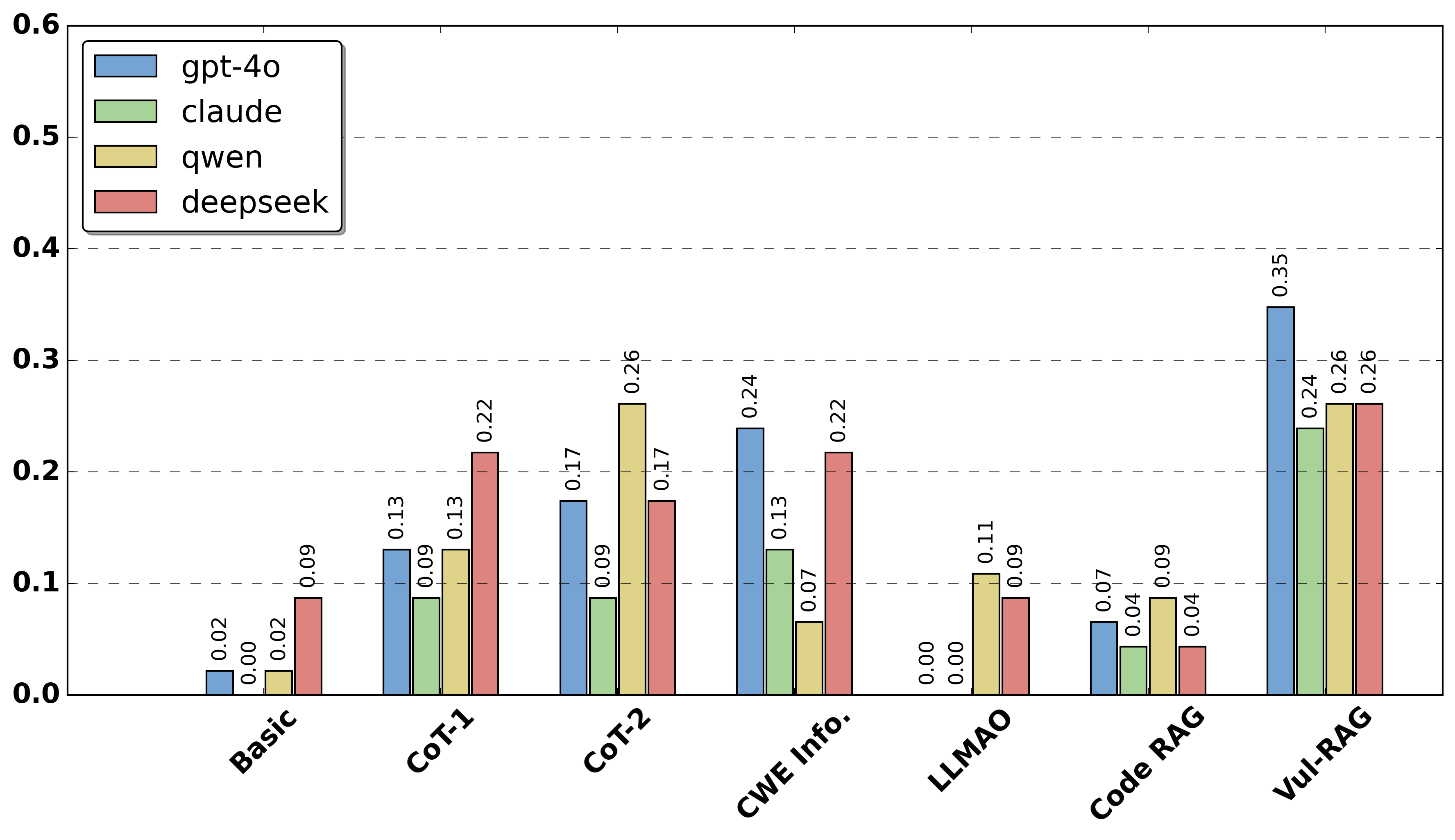}
        \label{fig:20}
    }
        \subfigure[CWE-200 Pair Accuracy Result]{
        \includegraphics[width=0.45\linewidth]{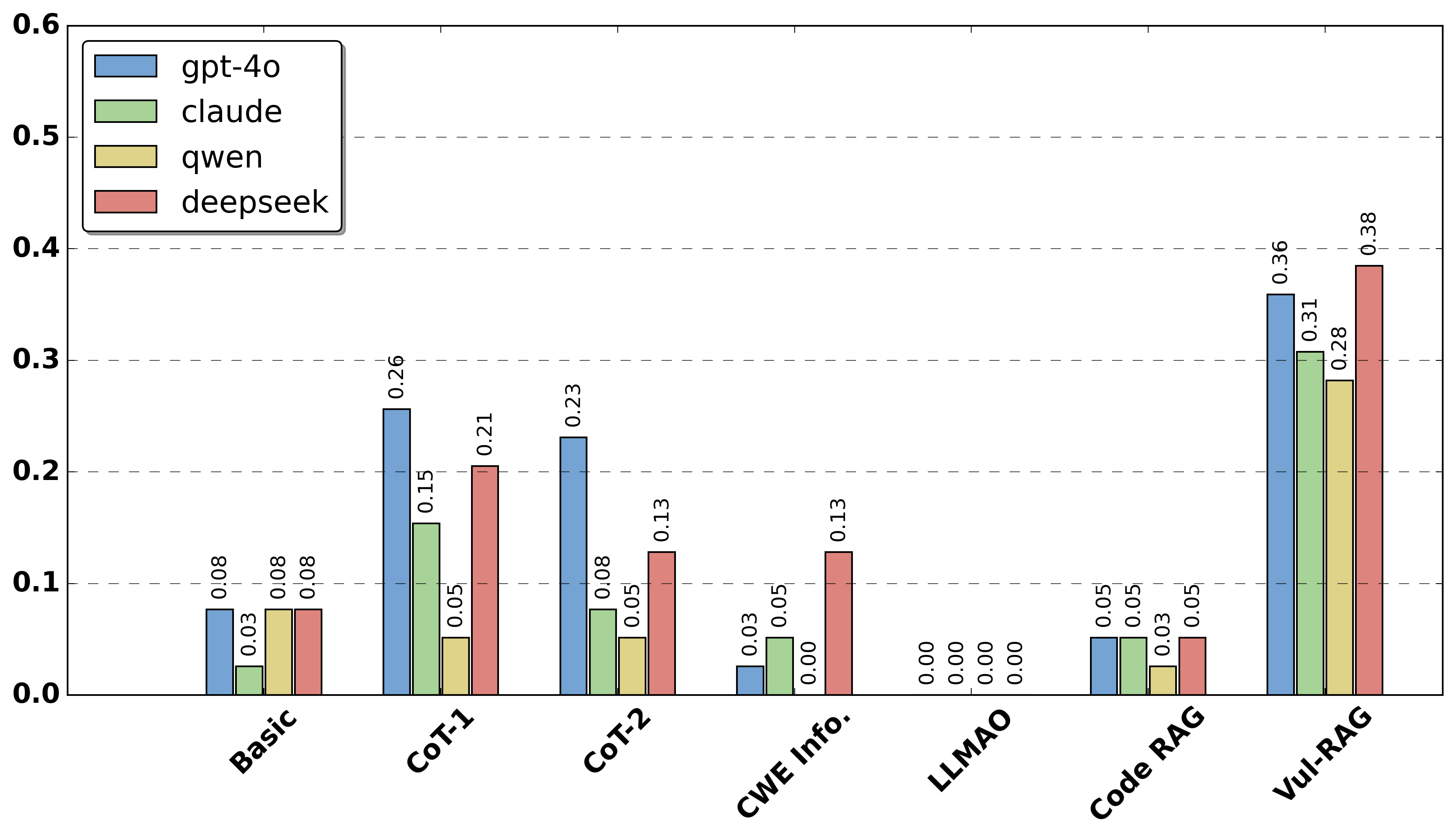}
        \label{fig:200}
    }
    \subfigure[CWE-125 Pair Accuracy Result]{
        \includegraphics[width=0.45\linewidth]{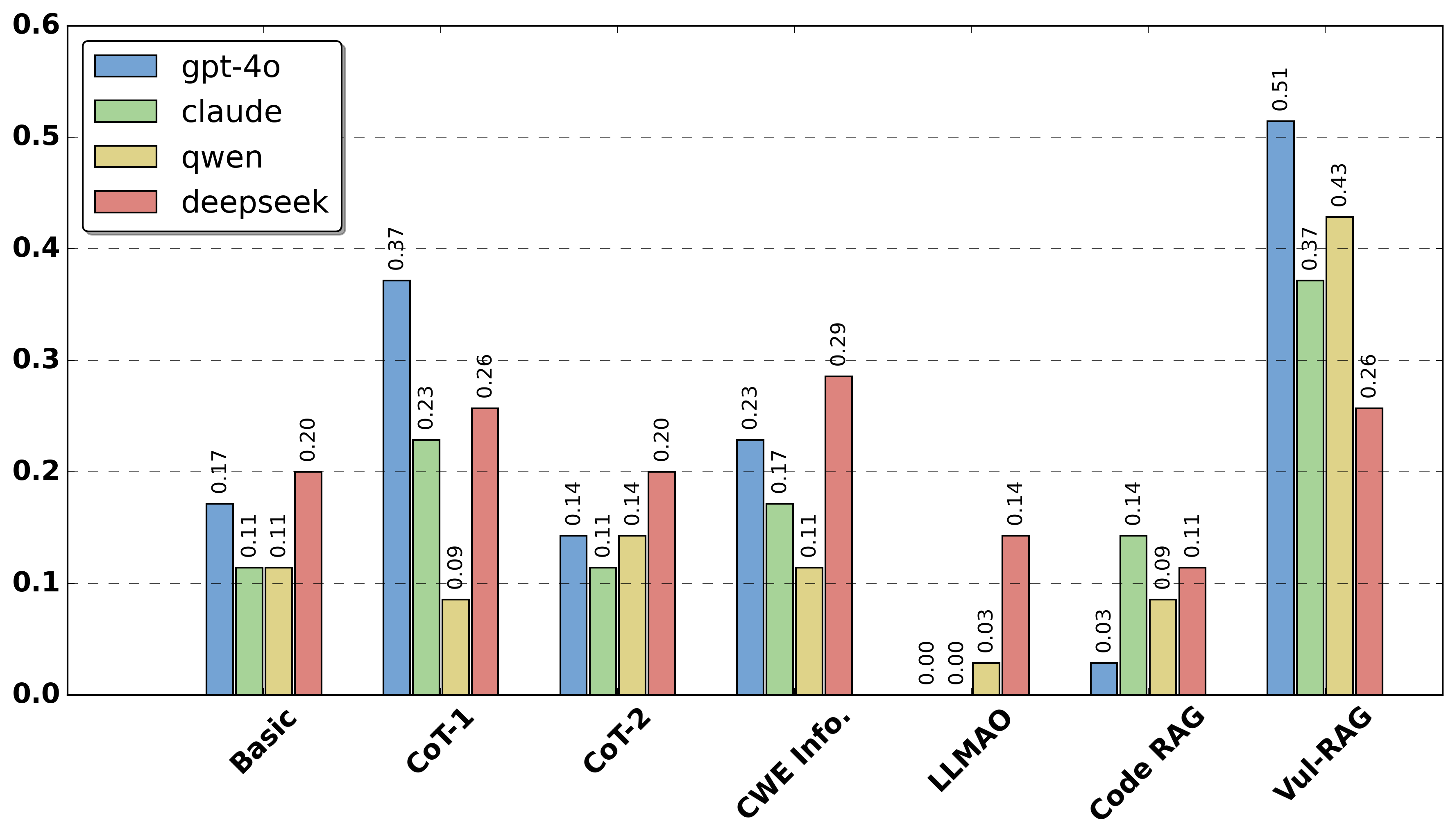}
        \label{fig:125}
    }
        \subfigure[CWE-264 Pair Accuracy Result]{
        \includegraphics[width=0.45\linewidth]{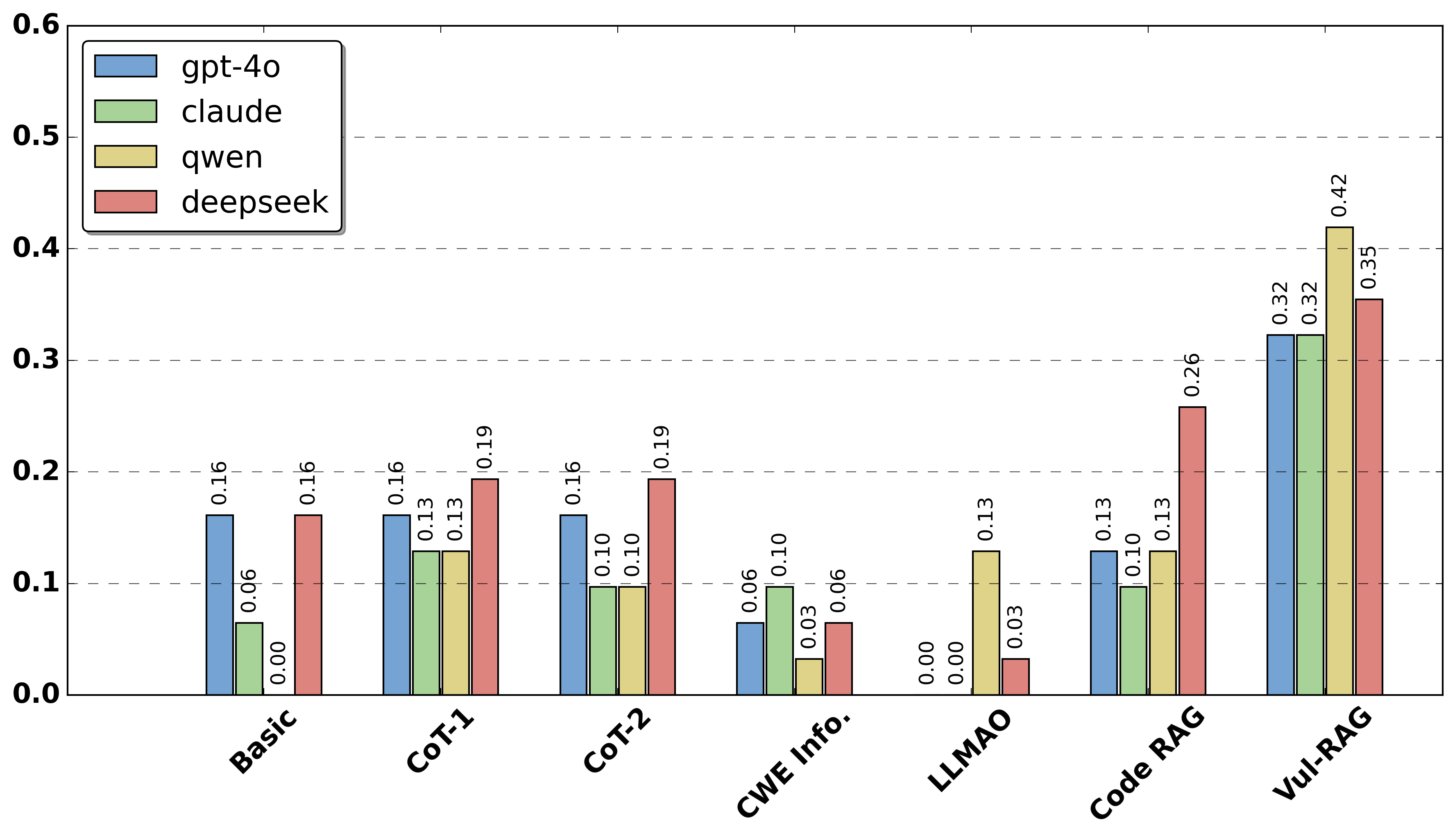}
        \label{fig:264}
    }
    \subfigure[CWE-401 Pair Accuracy Result]{
        \includegraphics[width=0.45\linewidth]{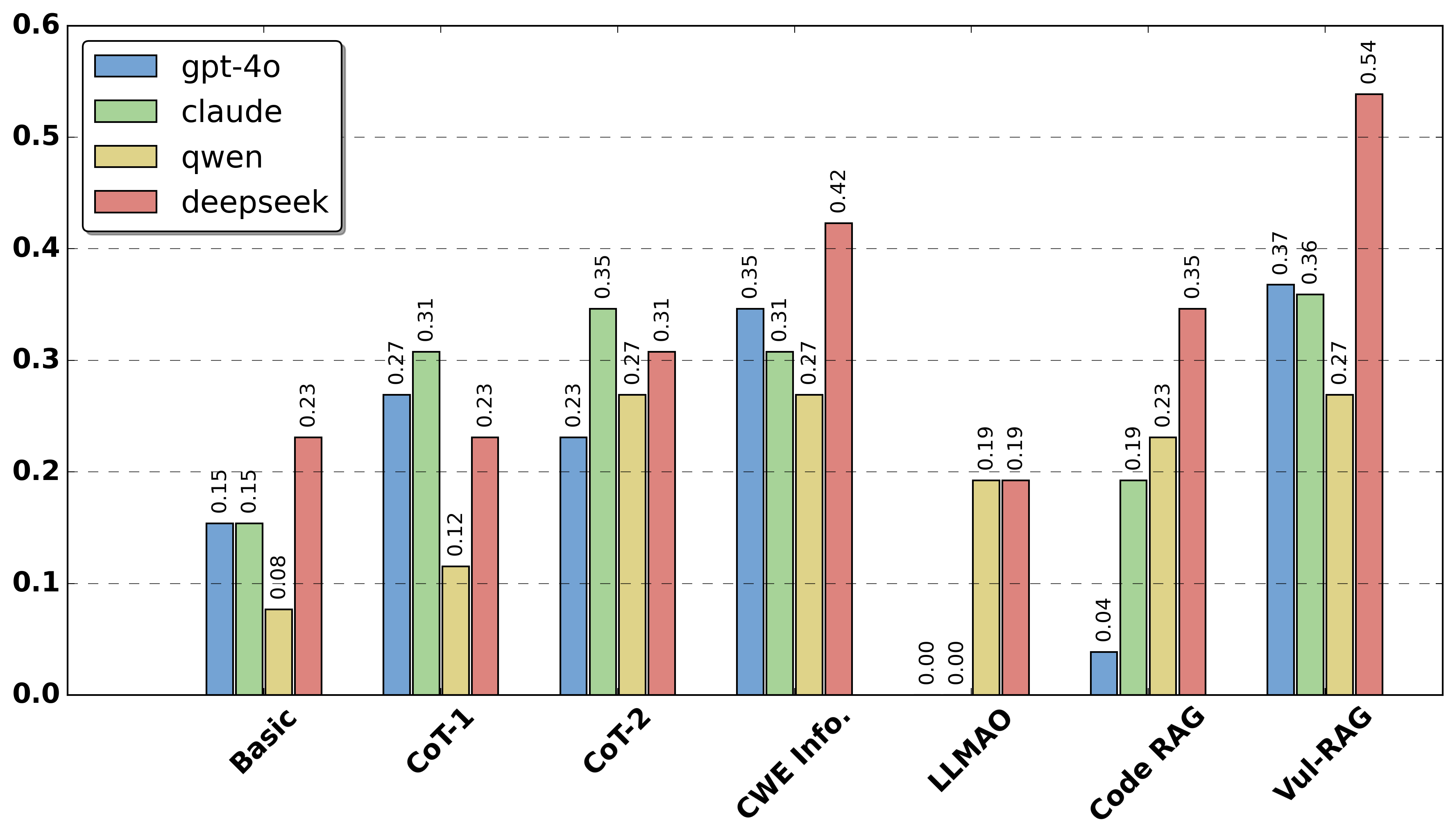}
        \label{fig:401}
    }
    \vspace{-3mm}     \caption{Comparison of performance for \app{} and Baselines}
    \label{fig:overall_performance}
\end{figure*}

Fig~\ref{fig:overall_performance} presents the performance of \app{} and all baselines across the 10 CWE categories.

\section{Case Study}
\label{ap:case}

\subsection{Case Study in Empirical Study}
\label{ap:case_study_rq}

\begin{figure*}[htb]
    \centering
    \subfigure[Example 1: Moving the location of a method invocation]{
        \includegraphics[width=0.8\linewidth]{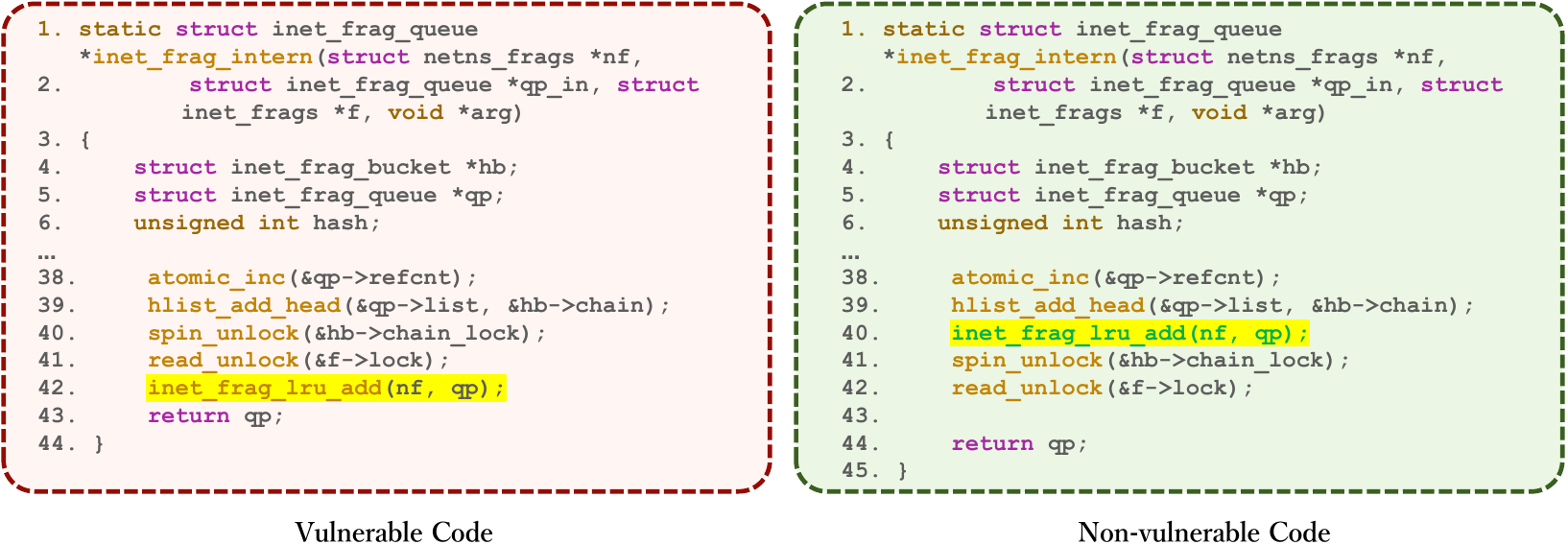}
        \label{fig:sim_code1}
    }
    \hfill
    \subfigure[Example 2: Adding a method invocation]{
        \includegraphics[width=0.8\linewidth]{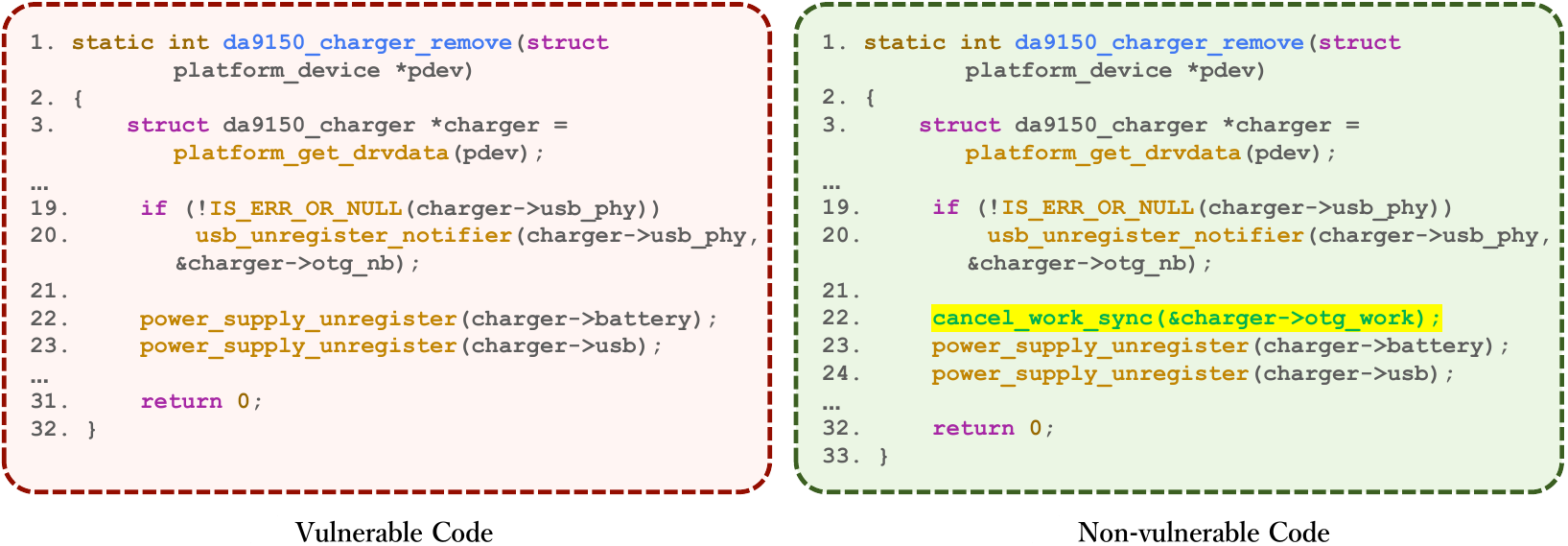}
        \label{fig:sim_code2}
    }
    \hfill
    \subfigure[Example 3: Adding a conditional check]{
        \includegraphics[width=0.8\linewidth]{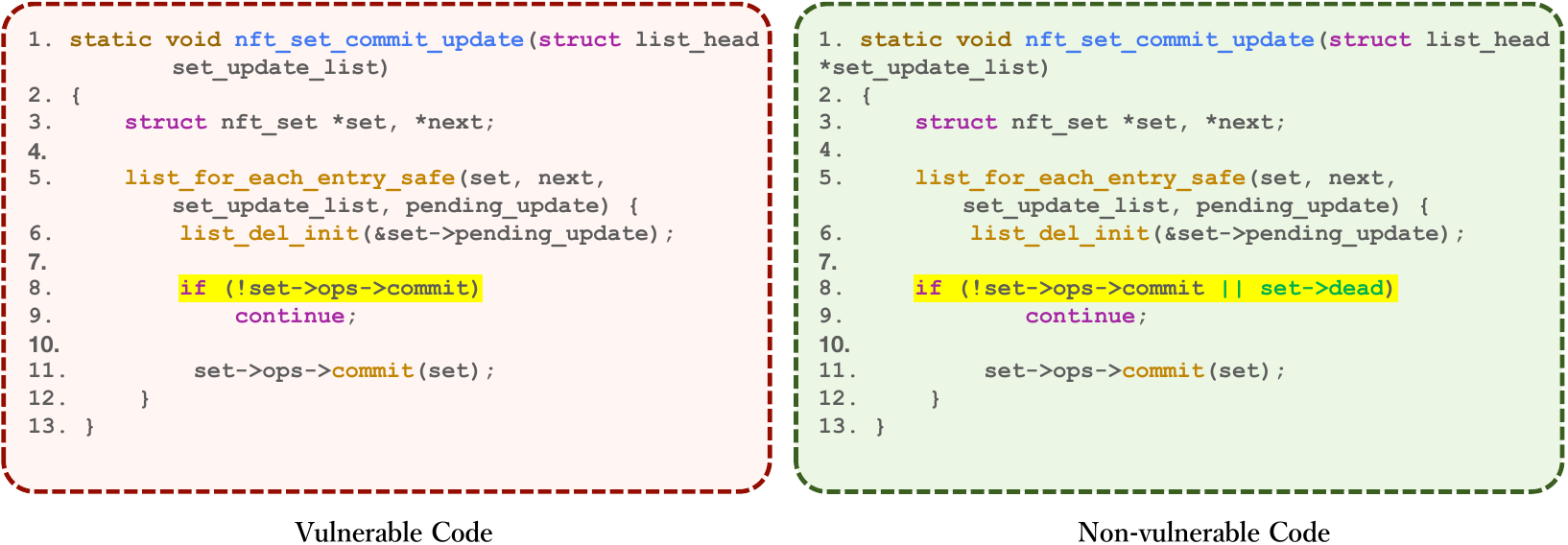}
        \label{fig:sim_code3}
    }
    \caption{Examples of vulnerable code and similar-but-benign patched code.}
    \label{fig:sim_code}
\end{figure*}

We sample and manually analyze pairs that all studied LLMs and advanced techniques in RQ1 and RQ2 fail to distinguish between vulnerable and patched code. 
Particularly, LLMs fail to distinguish the subtle textual difference between vulnerable code and patched code. Figure \ref{fig:sim_code} illustrates three specific examples, with the patch diffs highlighted in yellow.

\subsection{Case Study in Overall Improvements}

\label{ap:case_rq2}
\begin{figure*}[htb]
	\centering
	\includegraphics[width=0.9\linewidth]{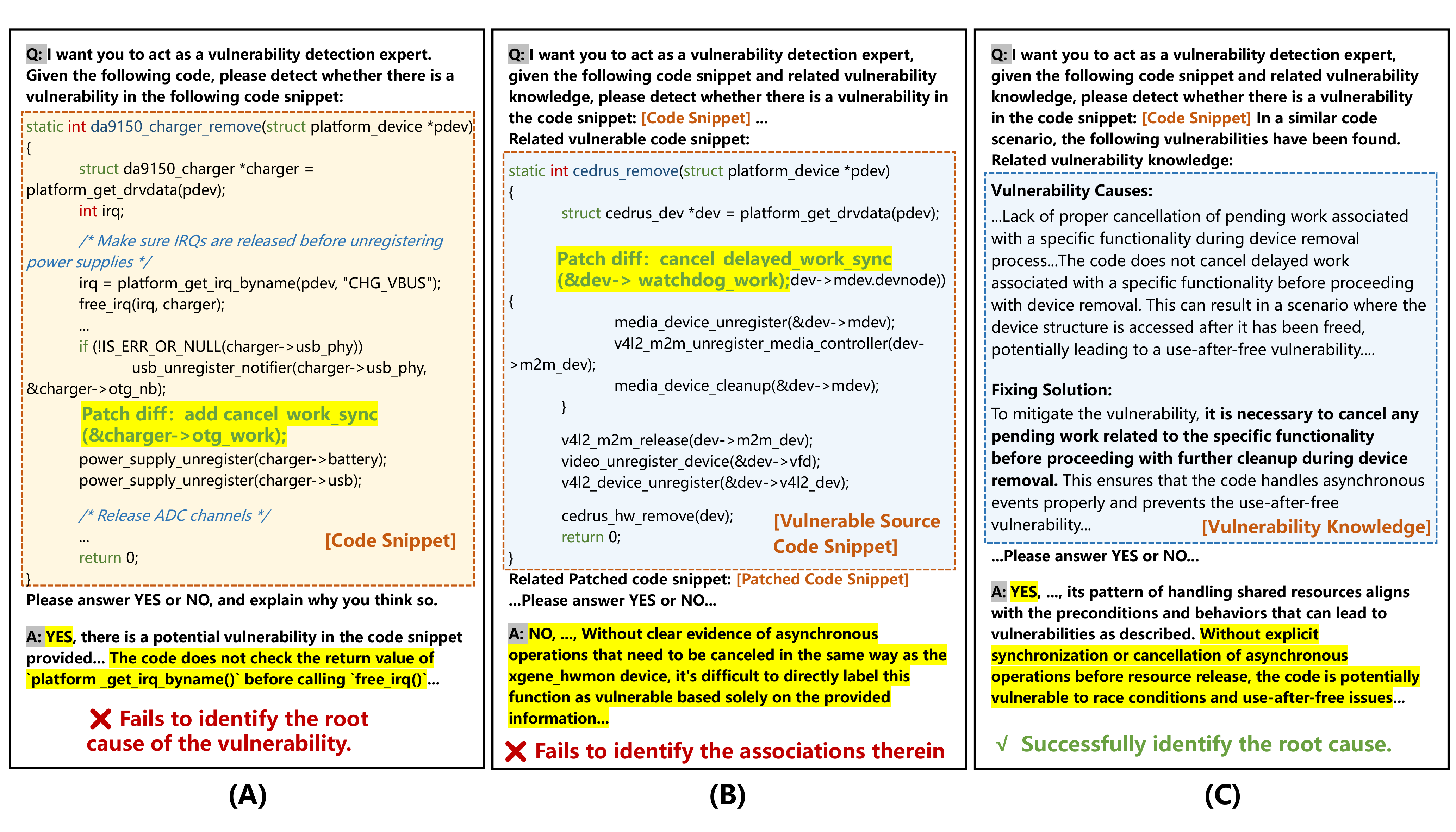}
	\caption{An example of vulnerability knowledge representation}
	\label{fig:mov_example}
        
\end{figure*}
\begin{figure*}[htb]
	\centering
	\includegraphics[width=0.9\linewidth]{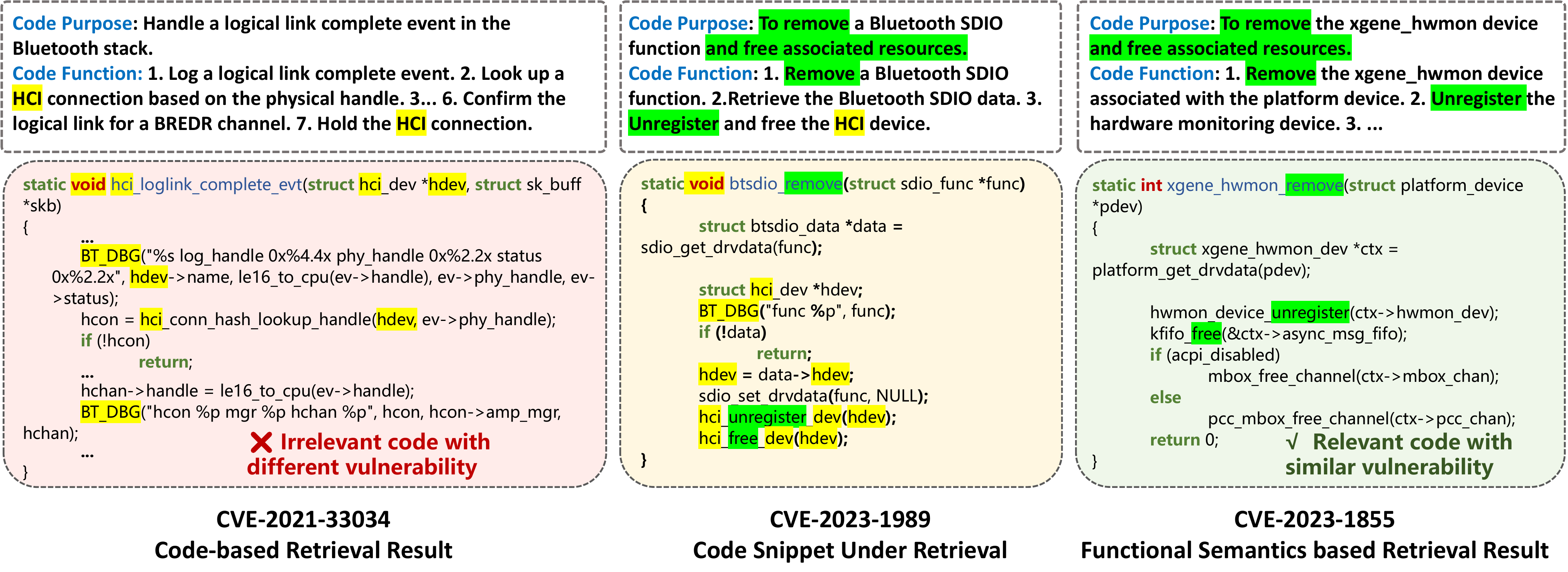}
	\caption{An example of knowledge retrieval strategy}
	\label{fig:retrieval_example}
    \vspace{2mm}
        
\end{figure*}

we use two examples that \app{} can successfully detect the vulnerability but code-level RAG cannot, to explain the superiority of \app{} in both knowledge representation and retrieval strategy.

\parabf{Knowledge Representation.}
Figure~\ref{fig:mov_example} illustrates an example to show the benefits of our knowledge representation comparing \app{} with basic LLM and code-level RAG baselines, all implemented on GPT-4. When detecting the given code from CVE-2023-30772, the basic GPT-4 fails to identify the real cause of the vulnerability (as shown in Figure~\ref{fig:mov_example} (A)).  GPT-4 incorrectly suggests that the absence of a return value check in \texttt{``platform\_get\_irq\_byname()''} could cause a vulnerability, whereas such a check is not required here. However, it overlooks the true issue, which is the improper handling of asynchronous events resulting in a race condition and subsequently a use-after-free vulnerability. This misunderstanding continues as GPT-4 detects the corresponding patched code, leading to false positives and affecting the pairwise accuracy. Enhancing GPT-4 with code-based RAG also fails to detect the vulnerability. As shown in Figure~\ref{fig:mov_example} (B), although the retrieved code pair contains a similar functional semantic and vulnerability cause, GPT-4 still struggles to associate the vulnerability knowledge implied in the retrieved source code with the target code under detection. In contrast, providing the distilled high-level vulnerability knowledge from our approach \app{}, GPT-4 not only successfully detects the vulnerability root cause in the vulnerable code but also accurately identifies the patched code (Figure~\ref{fig:mov_example} (C)). The comparison demonstrates the high-level vulnerability knowledge can effectively help LLMs understand the behavior of the vulnerable code, thereby improving the accuracy of vulnerability detection.

\parabf{Retrieval Strategy.}
Figure~\ref{fig:retrieval_example} compares the retrieving outcomes of code-based retrieval (\ie{} retrieving only by code snippet) and our retrieval strategy (\ie{} retrieving by both code snippet and extracted functional semantics) for the given code snippet. 
As shown in Figure~\ref{fig:retrieval_example}, when detecting a given code snippet from CVE-2023-1989, the code-based retrieval finds a code snippet (from CVE-2021-33034) that shares more operational resources with the target code (highlighted in yellow), but differ significantly in their functional semantics, leading to disparate root causes of vulnerabilities. In contrast, our retrieval strategy finds a code snippet (from CVE-2023-1855) that shares more semantic similarity with the target code (highlighted in green). Furthermore, they share an identical vulnerability root cause, which lies in the failure to adequately handle asynchronous events during the device removal process. This indicates that our retrieval strategy can help LLMs find code pairs with more similar vulnerability causes.

\subsection{Case Study of Previously-Unknown Vulnerability detected by \app{}}
\label{ap:unknown_bug}

\begin{figure*}[htb]
	\centering
	\includegraphics[width=0.9\linewidth]{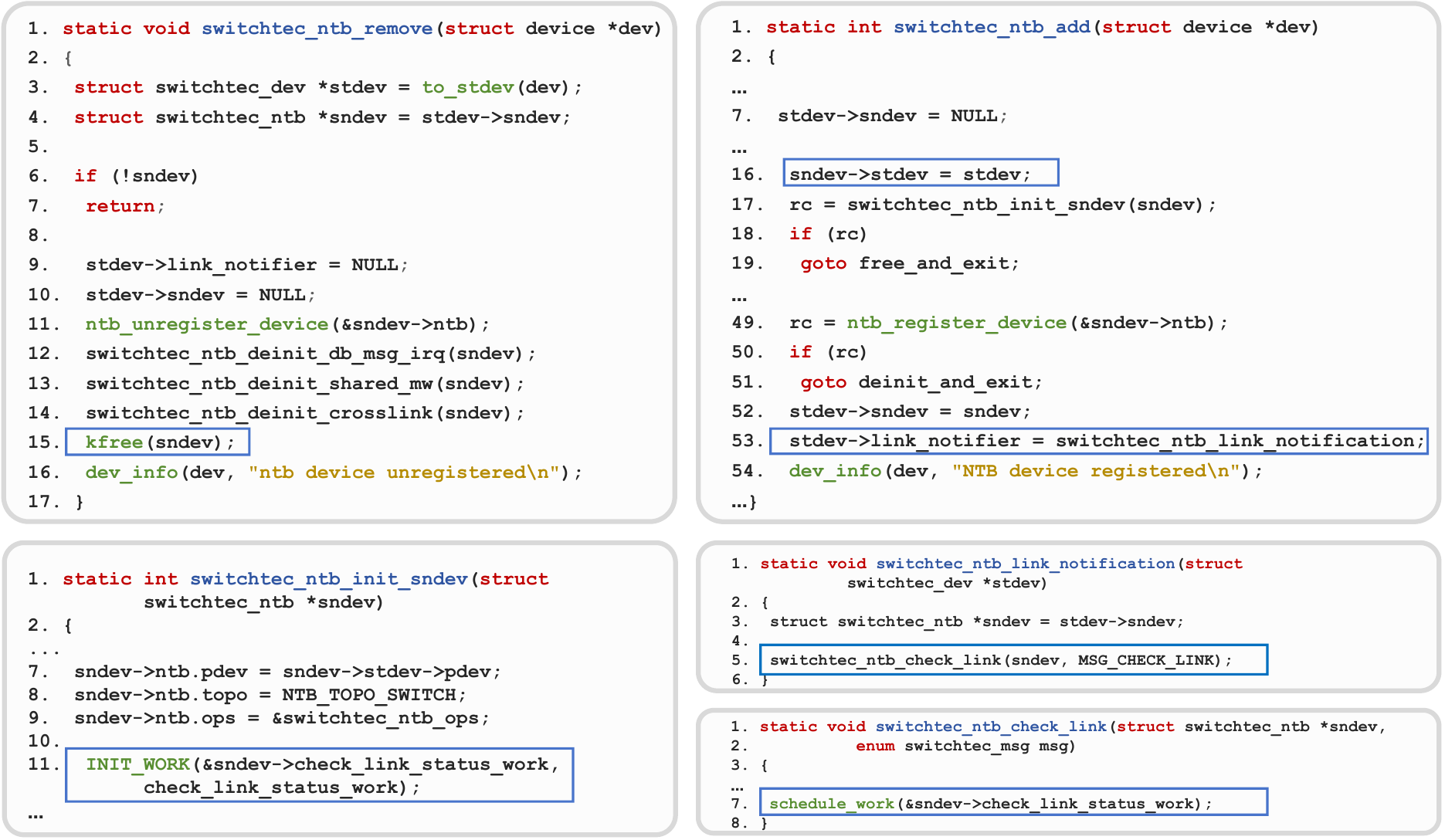}
	\caption{An example of a previously-unknown bug in Linux kernel reported by \app{}}
	\label{fig：real_bug_exm}
        
\end{figure*}

Figure~\ref{fig：real_bug_exm} shows a previously-unknown bug detected by \app{} in Linux kernel v6.9.6. This vulnerability is a use-after-free (UAF) caused by race condition found in the \textit{switchtec\_ntb\_remove} function located in \textit{drivers/ntb/hw/mscc/ntb\_hw\_switchtec.c} file. In \textit{switchtec\_ntb\_add} function, a call to switchtec\_ntb\_init\_sndev binds \&sndev->check\_link\_status\_work with check\_link\_status\_work. The \textit{switchtec\_ntb\_link\_notification} function may subsequently trigger the work by calling \textit{switchtec\_ ntb \_check\_link}. 
When \textit{switchtec\_ntb\_remove} is called during cleanup, it frees \textit{sndev} via kfree(sndev). If \textit{sndev} is accessed by CPU 1 via \textit{check\_link\_status\_work} after being freed by CPU 0, it could result in a use-after-free (UAF) vulnerability.
The vulnerability can be mitigated by ensuring that any pending work is canceled before the cleanup proceeds in \textit{switchtec\_ntb\_remove}, preventing access to memory that has been freed. Both the root cause and fixing solutions for this vulnerability align with those retrieved from CVE-2023-30772 in our constructed vulnerability knowledge base, demonstrating the scalability and effectiveness of the knowledge captured by \app{}.

\section{Usability for Developers}
\label{ap:rq4}

This section details the setup for our user study in investigating the quality of \app{} generated knowledge and whether the knowledge can help developers understand and check the vulnerabilities. 

\parabf{Tasks and Participants.} 
We select 10 cases from \ourbench{} for the user study. Specifically, we randomly select two cases from each of the five CWE categories \ourbench{}, including both true positive (\ie{} genuinely vulnerable code snippets) and false positive (\ie{} correct code snippets mistakenly predicted by \app{} as vulnerable) instances. To ensure a balanced evaluation, we randomly assign the two cases from each CWE category into two equal groups ($T_A$ and $T_B$), with each group comprising 5 cases. 
We invite 6 participants with 3-5 years c/c++ programming experience for the user study. We conduct a pre-experiment survey on their c/c++ programming expertise, based on which they are divided into two participant groups ($G_A$ and $G_B$) of similar expertise distribution.  Each participant are payed with 250\$ with the experiments. The procedure is approved by the Institutional Review Board (IRB) at our institution.

\parabf{Procedure.}
Each participant is tasked to identify whether the given code snippet is vulnerable. For comparison, participants are asked to identify vulnerability in two settings. (1) Basic setting: provided with the given code snippets and the detection labels generated by \app{}; (2) Knowledge-accompanied setting: provided with the given code snippets, the detection labels generated by \app{}, and the vulnerability knowledge generated by \app{}. In particular, the participants in $G_A$ are tasked to identify vulnerability in $T_A$ with the knowledge-accompanied setting, and to identify vulnerability in $T_B$ with the basic setting; conversely, the participants in $G_B$ are tasked to identify vulnerability in $T_A$ with the basic setting, and to identify vulnerability in $T_B$ with the knowledge-accompanied setting. 

In addition to recording the outputs (\ie{} vulnerable or not) of each participant, we further survey the participants on the helpfulness, preciseness, and generalizability of the vulnerability knowledge on a 4-point Likert scale~\cite{Likert1932} (\ie{} 1-disagree; 2-somewhat disagree; 3-somewhat agree; 4-agree). 

\begin{itemize}[leftmargin=10pt,itemsep=2pt,topsep=0pt,parsep=0pt]

\item \textbf{Helpfulness}: The vulnerability knowledge provided by \app{} is helpful in understanding the vulnerability and verifying detection labels.

\item \textbf{Preciseness}: The vulnerability knowledge offer precise and detailed descriptions of the vulnerability, avoiding overly generic narratives that do not adequately identify the root cause.

\item \textbf{Generalizability}: The vulnerability knowledge maintains a degree of general applicability, eschewing overly specific descriptions that diminish its broad utility (\eg{} narratives overly reliant on variable names from the source code).
\end{itemize}

In this evaluation, each human annotator is provided with the following instructions: \textit{``You will be presented with several C/C++ code snippets, each accompanied by a label indicating whether it is predicted as vulnerable or not. For each snippet, your task is to determine whether the code is truly vulnerable. For some snippets, you will also receive additional knowledge generated by the analysis tool to assist in your assessment. In these cases, you will be asked to evaluate the usefulness, clarity, and generalizability of the provided knowledge.''}

\section{Bad Case Analysis}
\label{ap:bad_case}

To understand the limitation of \app{}, we manually analyze the bad cases (\ie{} false negatives and false positives reported by \app{}). In particular, we include all 19 FN and 21 FP cases from CWE-119 for manual analysis. Table~\ref{table:rq4} summarizes the reasons and distributions.  In particular, the reasons for false negatives are  classified into three primary categories:

\begin{table}[htb]
	\centering
    \caption{FN/FP analysis in CWE-119}
    \label{table:rq4}
    \footnotesize
    \begin{adjustbox}{width=0.75\linewidth}
    \begin{tabular}{m{1cm}<{\centering}|m{7cm}<{\raggedright}|m{1cm}<{\centering}} \hline
    \textbf{Type} & \textbf{Reason} & \textbf{Number} \\  \hline
    \multirow{2}{*}{FN} & Inaccurate vulnerability knowledge descriptions. & 5 \\ \cline{2-3}
    & Unretrieved relevant vulnerability knowledge. & 2 \\ \cline{2-3}
    & Non-existent relevant vulnerability knowledge. & 12 \\ \cline{2-3} \hline 
    \multirow{2}{*}{FP} & Mismatched fixing solutions. & 11 \\ \cline{2-3}
    & Irrelevant vulnerability knowledge retrieval & 10 \\ \hline 
    \end{tabular}
    \end{adjustbox}
\end{table}

\begin{itemize}[leftmargin=10pt,itemsep=2pt,topsep=0pt,parsep=0pt]

\item \textbf{Inaccurate Vulnerability Knowledge Descriptions.} We observe that for 5 instances (26.3\%), \app{} successfully retrieves relevant vulnerability knowledge but fails to detect the vulnerability due to the imprecise knowledge descriptions. For example, given the vulnerable code snippet of CVE-2021-4204,  although \app{} successfully retrieves the relevant knowledge of the same CVE, it yields a false negative due to the vague descriptions of vulnerability knowledge (\ie{} only briefly mentioning ``\textit{lacks proper bounds checking}'' in the vulnerability cause and fixing solution description with explicitly stating what kind of bound checking should be performed). 

\item  \textbf{Unretrieved Relevant Vulnerability Knowledge.} We observe that for 2 cases (15.8\%) \app{} fails to retrieve relevant vulnerability knowledge, thus leading to false negatives. Although there are instances in the knowledge base that share the similar vulnerability root causes and fixing solutions of the given code, their functional semantics are significantly different. Therefore, \app{} fails to retrieve them from the knowledge base.

\item  \textbf{Non-existent Relevant Vulnerability Knowledge.} Based on our manual checking,  the 12 cases (63.2 \%) in this category are cased by the absence of relevant vulnerability knowledge in our knowledge base. Even though there are other vulnerable and patched code pairs of the same CVE, the vulnerability behaviors and fixing solutions are dissimilar, rendering these cases unsolvable with the current knowledge base. This limitation is inherent to the RAG-based framework. In future work, we will further extend the knowledge base by extracting more CVE information to mitigate this issue.
\end{itemize}

In addition, the reasons for false positive can be classified into the following two categories:

\begin{itemize}[leftmargin=10pt,itemsep=2pt,topsep=0pt,parsep=0pt]

\item \textbf{Mismatched Fixing Solutions.} There are 11 cases (52.4 \%) that although \app{} successfully retrieves relevant vulnerability knowledge, the code snippet is still considered vulnerable, as it is considered not applied to the fixing solution of the retrieved knowledge. This is because one vulnerability can be fixed by more than one alternative solution.

\item \textbf{Irrelevant Vulnerability Knowledge Retrieval.} There are 10 (47.6\%) false positives caused by \app{} retrieving irrelevant vulnerability knowledge. Based on our manual inspection, these incorrectly-retrieved knowledge descriptions often generally contain ``missing proper validation of specific values'', which is too general for GPT4 to precisely identify the vulnerability. 

\end{itemize}
\section{Limitations}
\label{sec:limit}
The incompleteness of the knowledge base can limit the performance of \app{} in practice. Given the diversity of vulnerabilities, it is possible that there is no relevant historical vulnerabilities for the code under detection, which is also a common pain spot for RAG techniques.
Therefore, we plan to open source our vulnerability knowledge base, which can be further continuously maintained and extended by the community together. 
Furthermore, although we evaluated four LLMs, including both open-source and closed-source models, the generalizability of our findings to other LLMs requires further investigation.


\end{document}